\documentclass[twocolumn]{aastex61}
\usepackage{amsmath}
\usepackage{enumerate}
\usepackage[shortlabels]{enumitem}
\setlist[enumerate]{nosep,nolistsep}
\setlist[itemize]{nosep,leftmargin=*}

\graphicspath{{Figures/}}

\newcommand{\blos}[1]{{$\Delta{B}_{{\rm LOS}}$#1}}
\newcommand{\bt}[1]{{$B_{{\rm LOS}}(t)$#1}}
\newcommand{\BLOS}[1]{{\Delta{B}_{{\rm LOS}}#1}}
\newcommand{\Mx}[1]{{Mx cm$^{-2}$#1}}
\newcommand{\dpil}[1]{{${D}_{{\rm PIL}}$#1}}

\shorttitle{Statistics of Photospheric Magnetic Field Changes During Flares}
\shortauthors{Castellanos Dur\'{a}n, Kleint \& Calvo-Mozo}
\received{}
\revised{}
\accepted{}
\submitjournal{ApJ}

\begin{document}

\title{A Statistical Study of Photospheric Magnetic Field Changes During 75 Solar Flares}

\correspondingauthor{J. S. Castellanos Dur\'an}
\email{castellanos@mps.mpg.de}

\author[0000-0003-4319-2009]{J. S. Castellanos Dur\'an}
\affiliation{Observatorio Astron\'omico Nacional, Universidad Nacional de Colombia, Carrera 45 No. 26 85, 11001 Bogot\'a, Colombia.}
\affiliation{Max Planck Institute for Solar System Research, Justus-von-Liebig-Weg 3, D-37077 G\"ottingen, Germany.}

\author[0000-0002-7791-3241]{L. Kleint}
\affiliation{University of Applied Sciences and Arts Northwestern
Switzerland, Bahnhofstrasse 6, 5210 Windisch, Switzerland.}
\affiliation{Kiepenheuer-Institut f\"ur Sonnenphysik (KIS), Sch\"oneckstrasse 6, D-79104 Freiburg, Germany.}

\author[0000-0002-5041-1743]{B. Calvo-Mozo}
\affiliation{Observatorio Astron\'omico Nacional, Universidad Nacional de Colombia, Carrera 45 No. 26 85, 11001 Bogot\'a, Colombia.}

\begin{abstract}
Abrupt and permanent changes of photospheric magnetic fields have been observed during solar flares. The changes seem to be linked to the reconfiguration of magnetic fields, but their origin is still unclear.
We carried out a statistical analysis of permanent line-of-sight  magnetic field (B$_{\rm LOS}$) changes during 18 X-, 37 M-, 19 C-, and 1 B-class flares using data from {\it Solar Dynamics Observatory}/Helioseismic and Magnetic Imager. We investigated the properties of permanent changes, such as frequency, areas, and locations. We detected changes of B$_{\rm LOS}$ in 59/75 flares. We find that strong flares are more likely to show changes, with all flares $\ge$ M1.6 exhibiting them. For weaker flares, permanent changes are observed in 6/17 C-flares. 34.3\% of the permanent changes occurred in the penumbra and 18.9\% in the umbra. Parts of the penumbra appeared or disappeared in 23/75 flares. The area where permanent changes occur is larger for stronger flares. Strong flares also show a larger change of flux, but there is no dependence of the magnetic flux change on the heliocentric angle. The mean rate of change of flare-related magnetic field changes is 20.7 Mx cm$^{-2}$ min$^{-1}$. The number of permanent changes decays exponentially with distance from the polarity inversion line. 
The frequency of the strength of permanent changes decreases exponentially, and permanent changes up to 750 Mx cm$^{-2}$ were observed. We conclude that permanent magnetic field changes are a common phenomenon during flares, and future studies will clarify their relation to accelerated electrons, white-light emission, and sunquakes to further investigate their origin. 

\end{abstract}

\section{Introduction}
Free energy is stored in magnetic fields, which makes them the crucial part in powering flares. \citet{Wang1992} and \citet{Wang1994} measured sudden increases of the magnetic shear along the polarity inversion line (PIL) during X-class flares, which can be interpreted as permanent changes of photospheric magnetic fields. Abrupt and co-temporal photospheric magnetic field changes are commonly observed during flares \citep[e.g.,][]{Cameron1999, Kosovichev2001} and the field strengths do not return to their original values {according to data} taken hours after the flares \citep[e.g.,][for {a} review see \citealt{Wang2015}]{Spirock2002}. A statistical analysis of permanent changes of line-of-sight (LOS) magnetic field strengths (\blos{}) using ground-based observations found typical values of 90 \Mx{} with rates of change up to 200 \Mx{} min$^{-1}$ \citep{Sudol2005}. The median change is higher for X- than it is for M-class flares, and the unsigned magnetic flux change was found to have a modest correlation to the {\it GOES} classes of the flares \citep{Petrie2010}. But the origins of \blos{}, frequency distributions, as well as their relationship with the energy of the flares are still not well understood.
 
The structure of active regions (ARs) can be disturbed after flares \citep[e.g.,][]{Wang2002}. Penumbral areas and pores may decay  \citep[e.g.,][]{Wang2004, Deng2005, Liu2005}, and appear \citep{Wang2002a} associated to \blos{} and flares. \blos{} tend to be primarily located in areas with strong magnetic fluxes and near the PIL \citep[e.g.,][]{Kosovichev2001,Burtseva2015}. {Vector magnetic field observations showed} that the transverse component increased around the PIL during 11 X-class flares \citep{Wang2010}, and the field became more parallel to the flaring PIL \citep{Petrie2013}.

\citet{Hudson2000} suggested that when the coronal magnetic energy is released during flares, there should be a subsequent {\it implosion} and field lines may contract due to a reduction in the magnetic pressure. The contracted field lines overall become more horizontal after the flare, and a reaction may be generated as consequence of momentum conservation \citep{Hudson2008, Fisher2012}. Measurements of the magnetic field changes during flares, and extrapolations of nonlinear force-free coronal fields have been interpreted consistently with the coronal implosions picture in multiple studies \citep[e.g.,][]{Petrie2010, Wang2010, Gosain2012, Wang2012, Liu2012, Liu2014, Petrie2012, Petrie2016, Sun2012}. However, new observations showed a temporal and spatial incoherence between photospheric and chromospheric permanent changes of the LOS magnetic field \citep{Kleint2017}.  Furthermore, the observed tilt angles in the chromosphere did not support decreasing loop sizes after the flare, but rather the possibility of either increasing loop sizes or, more likely, the possibility of untwisting into an energetically favorable state \citep{Kleint2017}. 

In this paper, we carry out a statistical analysis of LOS permanent changes of the magnetic field during 75 flares to investigate their frequency, their dependence on flare energy, and their distance to the PIL. This is the first statistical study of magnetic field changes that takes into account a large range of flare energies, of flare locations, and 47 different ARs. 

\section{Observations and data reduction} \label{sec:obs}
\begin{figure}[tbh]
 \begin{center}
 \includegraphics[width=.49\textwidth]{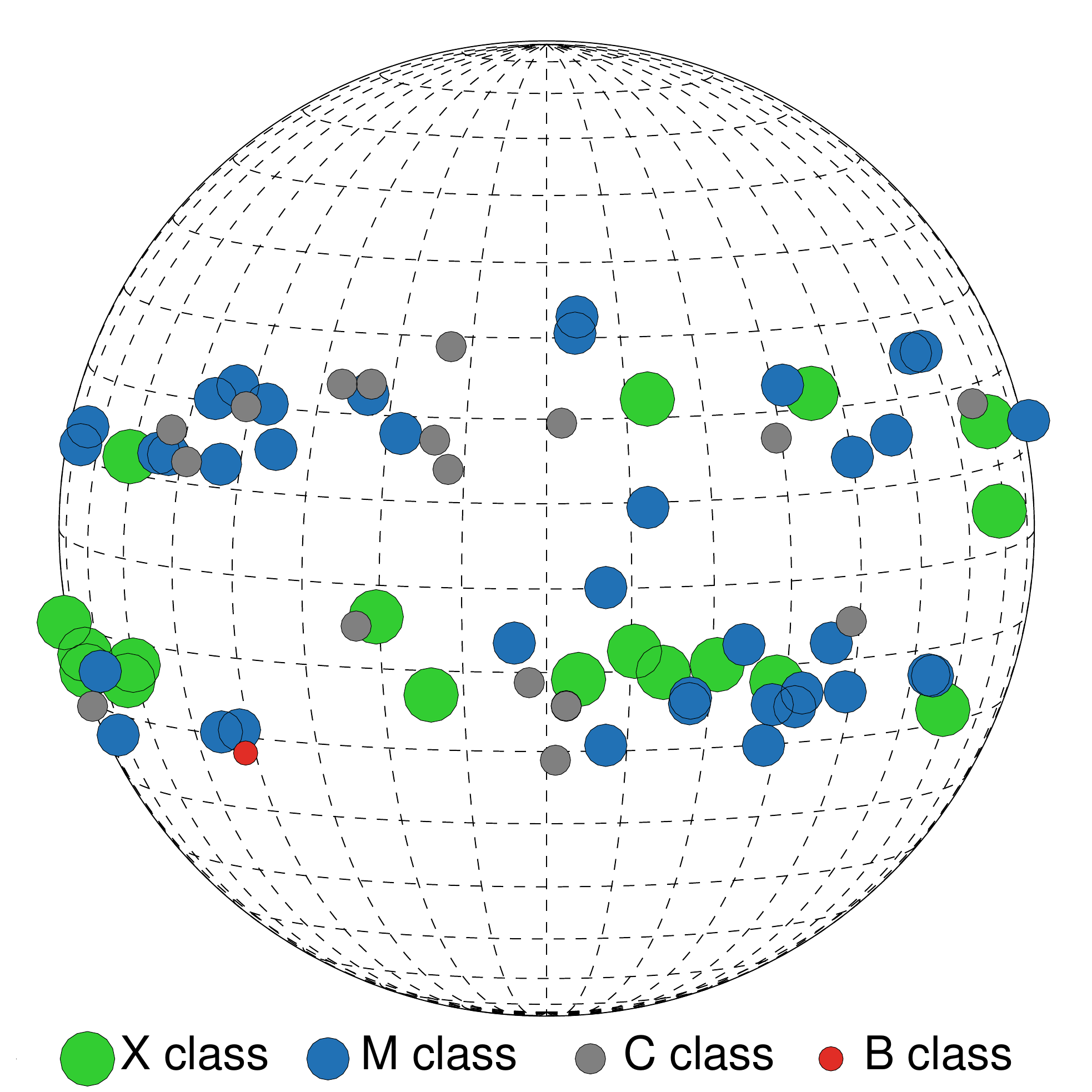}
 \caption{Distribution of the flare sample on the solar disk. Color-coded circles show the flare locations, where more energetic flares are associated with larger circles. Colors represent X- (green), M- (blue), C- (gray), and B-class (red) flares. Solar longitudes and latitudes are drawn with dashed lines every $10^{\circ}$.\label{fig:flareloc}}
 \end{center}
 \end{figure}

To perform a statistical analysis of \blos{}, we selected 75 flares with a large energy range and different locations on the solar disk. The energies ranged from 18 X-, 37 M-, and 19 C-, to 1 B-class events, which occurred from 2010 October to 2015 March. The flare locations, ranging from $\mu=0.27 - 0.97$, where $\mu$ denotes the cosine of the heliocentric angle ($\mu=\cos \theta$), are illustrated in figure \ref{fig:flareloc}. The sizes and colors of the circles represent {\it GOES} classes. Table \ref{tab_sample} {at the end of the manuscript} lists details of all flares used for this study. 

To study the \blos{} during flares, we used data from the Helioseismic and Magnetic Imager \cite[HMI,][]{Scherrer2012} onboard the {\it Solar Dynamics Observatory} \citep[{\it SDO},][]{Pesnell2012}. We analyzed HMI magnetograms ({\it hmi.M\_45s}) and continuum intensity images ({\it hmi.Ic\_45s})  with a cadence of 45 s and plate scale of 0\farcs504 pixel$^{-1}$.  HMI samples the spectral region around the \ion{Fe}{1} 6173.3 \AA\ absorption line at six wavelength points with a bandwidth of 76 m\AA. It is known that HMI magnetograms underestimate strong magnetic fields \citep{Couvidat2012a}, for example, when they are compared with LOS magnetic fields retrieved from Milne-Eddington inversions \citep{Hoeksema2014}. Some flares show transient changes in magnetograms, which may appear as a sudden apparent reversal of magnetic polarities. Some observations and calculations explained the transient as a result of Doppler shifts and changes in the shape of the spectral line \citep[e.g.][]{Patterson1984, Ding2002, Qiu2003}, and some consider them real \citep{Harker2013}. For our analysis, we therefore ignored the five points temporally closest to the {\it GOES} peak time of each flare, including the peak itself.

\begin{figure*}[htbp]
\begin{center}
\includegraphics[width=1.\textwidth]{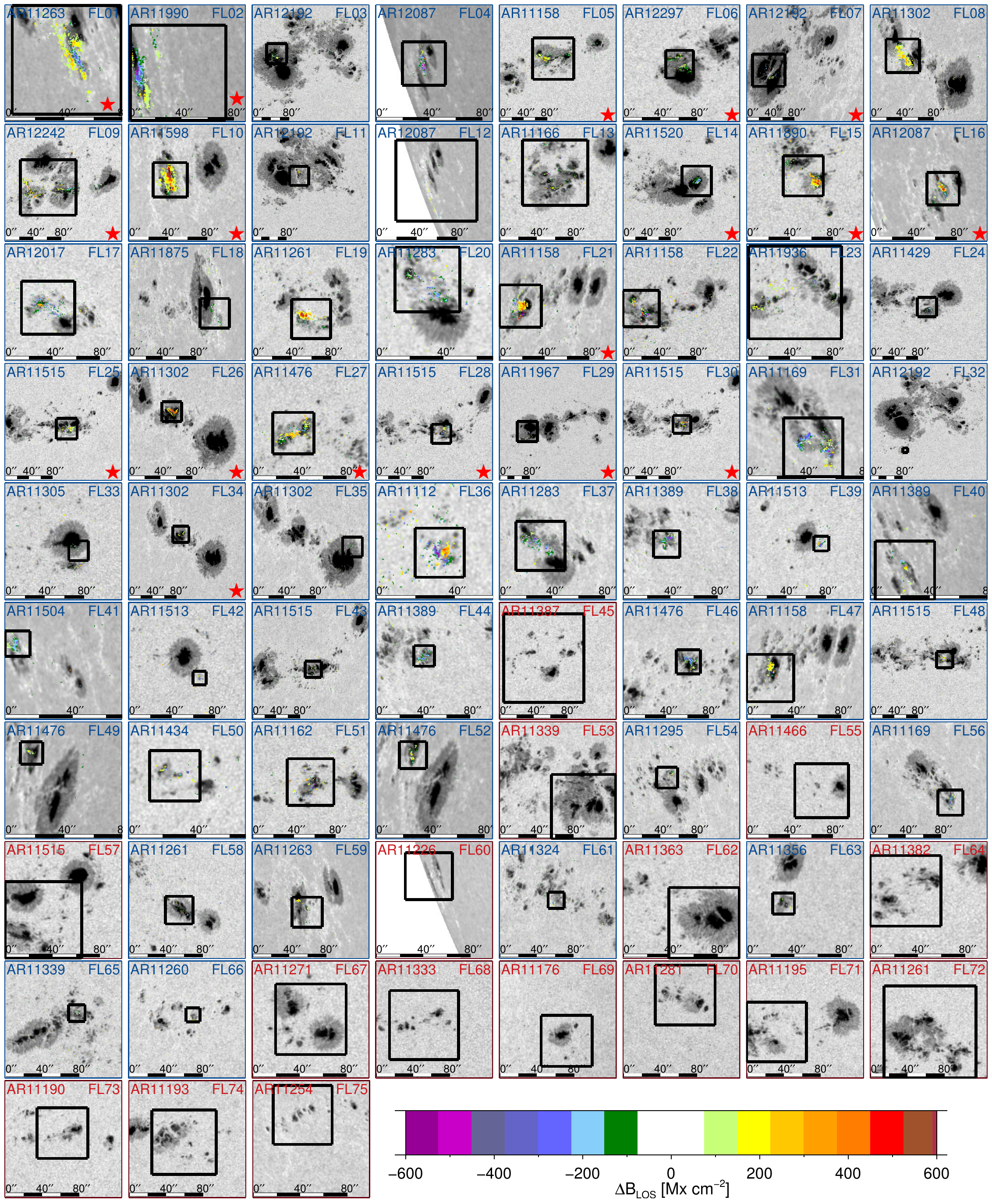} 
\caption{HMI intensity images showing the flare sample, ordered by decreasing flare energy. Color-coded pixels indicate locations with \blos{}, which are clipped at $\pm$600 \Mx{} according to the color bar on the bottom. Blue axes and labels are used for flares where \blos{} were found. Red axes and labels represent flares without \blos{}. Image scales are shown by the dashed bar with a length of 80\arcsec{} in each image. Black boxes in the intensity images indicate the FoVs in figure~\protect\ref{fig:mag_kernelchanges_zoom}. {Red stars denote the 18 flares that, in addition to classical stepwise changes, also showed other types of temporal evolutions of \bt{} (see section \ref{sec:type}).}  \label{fig:int_kernelchanges_full}}
\end{center}
\end{figure*}

The magnetic structures of the AR, the chromospheric and coronal emission change for each flare. A different field of view (FoV) is therefore required for each event. We took the location of each flare from the {\it RHESSI}, AIA, or {\it GOES} catalogs to center the initial FoVs. We then used images from the Atmospheric Imaging Assembly \citep[AIA,][]{Lemen2012} to adapt the FoVs to enclose the full flare manually. AIA channels have an average cadence of 12 s and plate scale of 0\farcs6 pixel$^{-1}$. The FoVs in our sample range from 80\arcsec$\times$80\arcsec{} to 300\arcsec$\times$300\arcsec{}. Figure \ref{fig:int_kernelchanges_full} shows HMI continuum intensity images of the FoVs at the time of the {\it GOES} peak of each flare.

To apply a standardized data-reduction process, we developed a pipeline that only requires the {\it GOES} times, the coordinates of the flares, and the size of the FoV as input. The pipeline automatically downloads the data and organizes them; aligns and crops the HMI images; and fits a function that determines the \blos{} from the temporal evolution of the magnetic field \bt{} of each pixel (see next section). The main modules of the pipeline are described in the following section. We use units of \Mx{} (flux density) instead of Gauss  (field strength) to account for unresolved magnetic fields due to the spatial resolution.

\newpage 
\subsection{Locations and Amplitudes of \blos{}}\label{sec:locationblos}
We fitted  \bt{} of each pixel with a stepwise function \citep{Sudol2005} given by
\begin{equation}\label{eq:sudol}
{\rm B}_{{\rm LOS}}(t) = a + bt + c\left\{ 1 +\frac{2}{\pi}\tan^{-1}[n(t-t_0)]\right \},
\end{equation}
where $ a $, and $ b $ describe a linear evolution of the background field {with} time $t$. The parameter $ c $ is the half-size of the step, $ t_0 $ is the midpoint of the step, and $ n^{-1} $ is the duration of the stepwise change. The stepwise function was fitted with the nonlinear least square package MPFIT in IDL \citep{Markwardt2009}. 

To study the temporal evolution of the \blos{}, we need a trade-off between a short time range to exclude the evolution of the AR and a long enough time to observe permanent changes. We tested fitting six different time ranges centered at the {\it GOES} peak time of each flare: 480, 240, 120, 90, 60, and 30 minutes. Fits of time ranges $\ge2$ h showed the evolution of the AR, for example, formation and disappearance of pores or penumbra and did not yield flare-related magnetic field changes reliably.
Time ranges of 30 and 60 minutes were too short to observe certain slow permanent changes in long-duration flares. Therefore, we selected the 90-minute interval{, even though former studies sometimes considered longer intervals \citep[e.g. 4 hr in][]{Sudol2005,Petrie2010}}.

Previous studies used the parameter $2c$ from the stepwise function to measure the amplitude of the \blos{} \citep[e.g.,][]{Sudol2005,Petrie2010}, {thus assuming that the background field is fully described by a+bt and continues with the same slope throughout the flare. We sometimes found changes in the background field evolution at flare time, and therefore we decided to drop the assumption that the field continues to evolve linearly during the flare. We therefore estimate the step size more conservatively, which leads to differences to 2$c$ especially when the slope of the background field evolution is large.} An example is shown in the appendix. We used three methods to calculate and correct the retrieved step size. (1) We took the difference between the maximum and the minimum values of \bt{} before and after the flare. (2) $\Delta {\rm B}_{{\rm LOS}} = 2c$ \citep{Sudol2005}. (3) We performed a geometrical correction to measure the size of the step, which takes into account the duration of the step $n^{-1}$ and the slope of the background $b$ from equation \ref{eq:sudol}. We calculated two parallel lines intercepting the stepwise function at the start ($t_s=t_0 - \pi n^{-1}$) and {at the} end ($t_e=t_0 + \pi n^{-1}$) times of the step. The distance between the two parallel lines is the size of the change, which is given by
$\Delta {\rm B}_{{\rm LOS}} = {\rm B}_{{\rm LOS}}(t_e) - {\rm B}_{{\rm LOS}}(t_s) - 2\pi bn^{-1}.$
The smallest value of the methods (1)-(3) was selected as the amplitude of the magnetic field change \blos{}. 

The stepwise function (Equation \ref{eq:sudol}) sometimes fitted transient changes, or failed in cases where two steps were observed. Therefore, we manually examined all profiles with \blos{} $\ge$ 80 Mx cm$^{-2}$ and discarded pixels without stepwise and permanent changes. Of the $6\times10^6$ pixels in our full sample, $1.2\times10^6$  yielded fits with \blos{}$\ge$80 \Mx{}. Our manual verification of these $1.2\times10^6$ pixels was performed twice and only differed by 5\% for the two runs. We found 27,153 {permanent changes of the evolution of \bt{}}. {For 91\% of these changes, Eq.~\ref{eq:sudol} yielded good fits based on $\chi^2$, the goodness-of-fit value.  For the remaining 9\%, either the step size was still reported well by one of the three methods even though not the whole \bt{} was fitted well, or the step size was determined manually.}

\subsection{{Evolution of the Magnetic Field During Flares}}\label{sec:type}

\begin{figure}[b]
\begin{center}
\includegraphics[width=.47\textwidth]{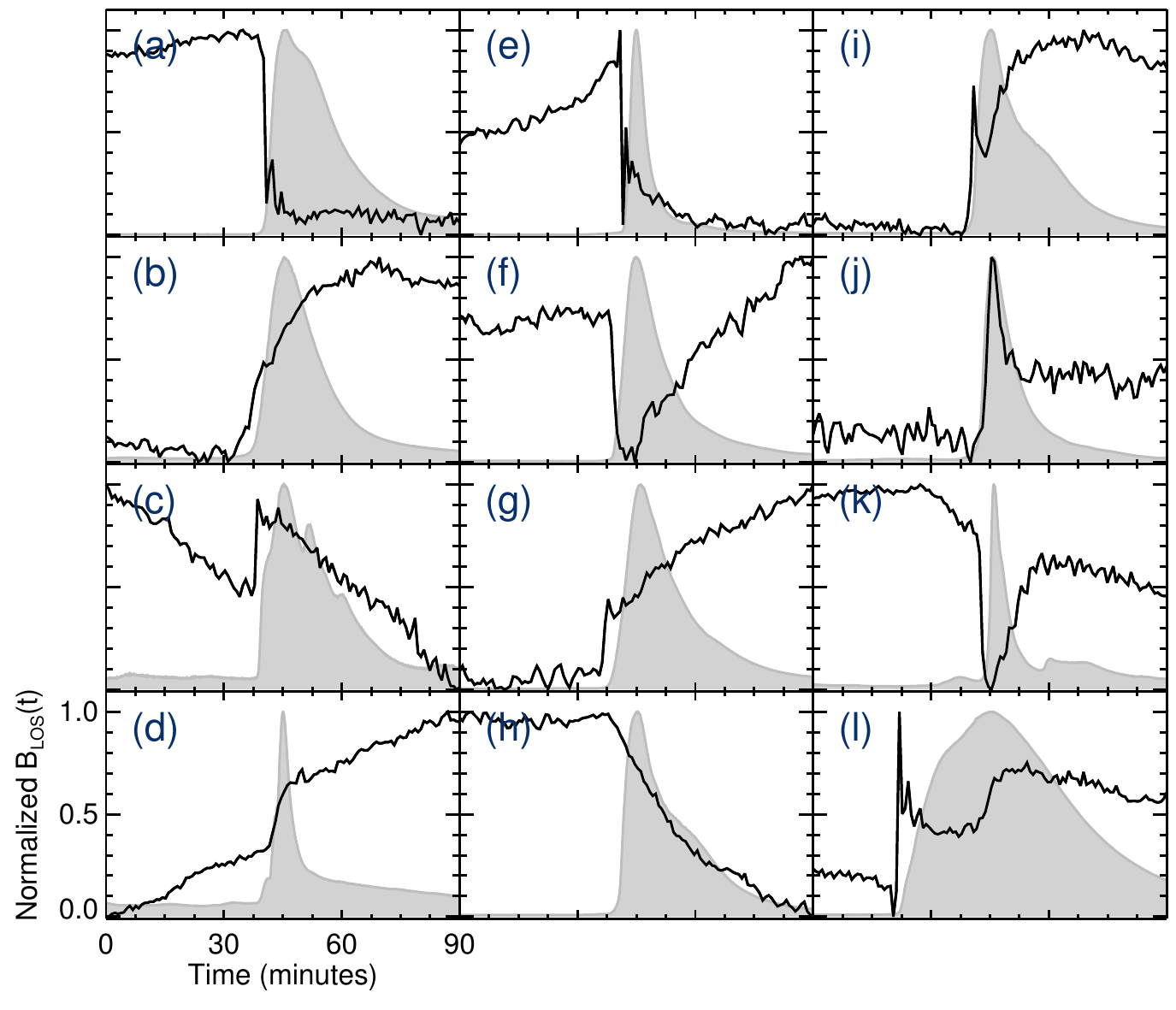} 
\end{center}
\caption{Twelve {types of temporal evolutions of \bt{}} classified manually by their morphology. The $x$-axis ranges from 0 to 90 minutes centered on the maximum of the flare. The $y$-axis shows the normalized \bt{}, and gray shades are the normalized {\it GOES} 1 - 8 \AA{} curves for the flares. Note. Seven of the present types of changes were previously reported in: (a) [2-6,8-10,13-15], (b) [1-5,7,8,11-20], (c) [2,12,15], (e) [7], (i) [15], (j) [4,21,22], and (k) [2-4,6,7,9,21]. References: (1) \citealt{Wang1992}, \citealt{Wang1994}; (2) \citealt{Sudol2005}; (3) \citealt{Petrie2010}; (4) \citealt{Chen2007}; (5) \citealt{Cliver2012}; (6) \citealt{Johnstone2012}; (7) \citealt{Kosovichev2001}; (8) \citealt{Petrie2012}; (9) \citealt{Liu2005}; (10) \citealt{Liu2014}; (11) \citealt{Petrie2013}; (12) \citealt{Yurchyshyn2004}; (13) \citealt{Wang2012}; (14) \citealt{Wang2010}; (15) \citealt{Meunier2003}; (16) \citealt{Deng2005}; (17) \citealt{Wang2002}; (18) \citealt{Wang2002a}; (19) \citealt{Wang2012}; (20) \citealt{Wang2010}; (21) \citealt{Burtseva2015}; (22) \citealt{Gosain2012}.\label{fig:classes}}
\end{figure}

The durations and behaviors of the permanent changes differ \citep[e.g.,][]{Sudol2005, Petrie2010, Cliver2012, Burtseva2015}. Not all changes in the evolution of \bt{} found in our sample can be described well with a simple stepwise model, as given by equation~\ref{eq:sudol} (e.g. figure \ref{fig:classes}, panels f, g, and l). 

We visually found 12 different types of {changes of \bt{}} in our sample, {not all of which are considered permanent and stepwise}. Figure \ref{fig:classes} shows examples that describe their morphological characteristics; gray shades in the background are the {\it GOES} curves of the example flares. We sorted these types of {\bt{}} into categories based on their visual appearance.  The characteristics of the types are as follows.

(1) The permanent change has either a long duration ($>$3 minutes, Figure~\ref{fig:classes}(a)) or a short duration ($<$3 minutes, Figure~\ref{fig:classes}(b)). 
(2) The background of \bt{} never (Figure~\ref{fig:classes}(a)-(b)), sometimes (Figure~\ref{fig:classes}(e)-(h)), or always (Figure~\ref{fig:classes}(c)-(d)) shows a slope before and after the permanent change in our considered time range.
(3) There are {no} changes in the slope of the background of \bt{} before and after the permanent change (e.g., Figure~\ref{fig:classes}(a)-(d)), {or} the magnitude and/or sign of the slope may change (Figure~\ref{fig:classes}(f)-(h)). A slope $\neq0$ may imply flux emergence or cancellation, though the investigation of emergence/cancellation of magnetic flux is out of the scope of the present work \citep[a previous study of this topic can be found in][]{Wang2004a, Burtseva2013}. 
(4) A sudden peak appears near the maximum of the flare (Figure~\ref{fig:classes}(i)-(k)), which may imply a magnetic transient {(such points were omitted before fitting Equation \ref{eq:sudol})}.
(5) Number of stepwise changes: zero (Figure~\ref{fig:classes}(f)), one (Figure~\ref{fig:classes}(a)-(k)), and two (Figure~\ref{fig:classes}(l)). The note in Figure \ref{fig:classes} lists examples of previous studies, where a subset of the {types of changes} were reported. {All of the types of changes seem flare-related, but when the slope differs before and after the change, the background field evolution may contribute to the reported step size. We therefore excluded changes of types f (2.1 \%), g (0.5 \%), h (0.5 \%), and k (2.4 \%) from our final sample. This reduced the 27,153 detected changes by 1495 to 25,658 stepwise changes that all fall into types a, b, c, d, e, i, j, and l.}

\begin{figure*}[tbph]
\includegraphics[width=.99\textwidth]{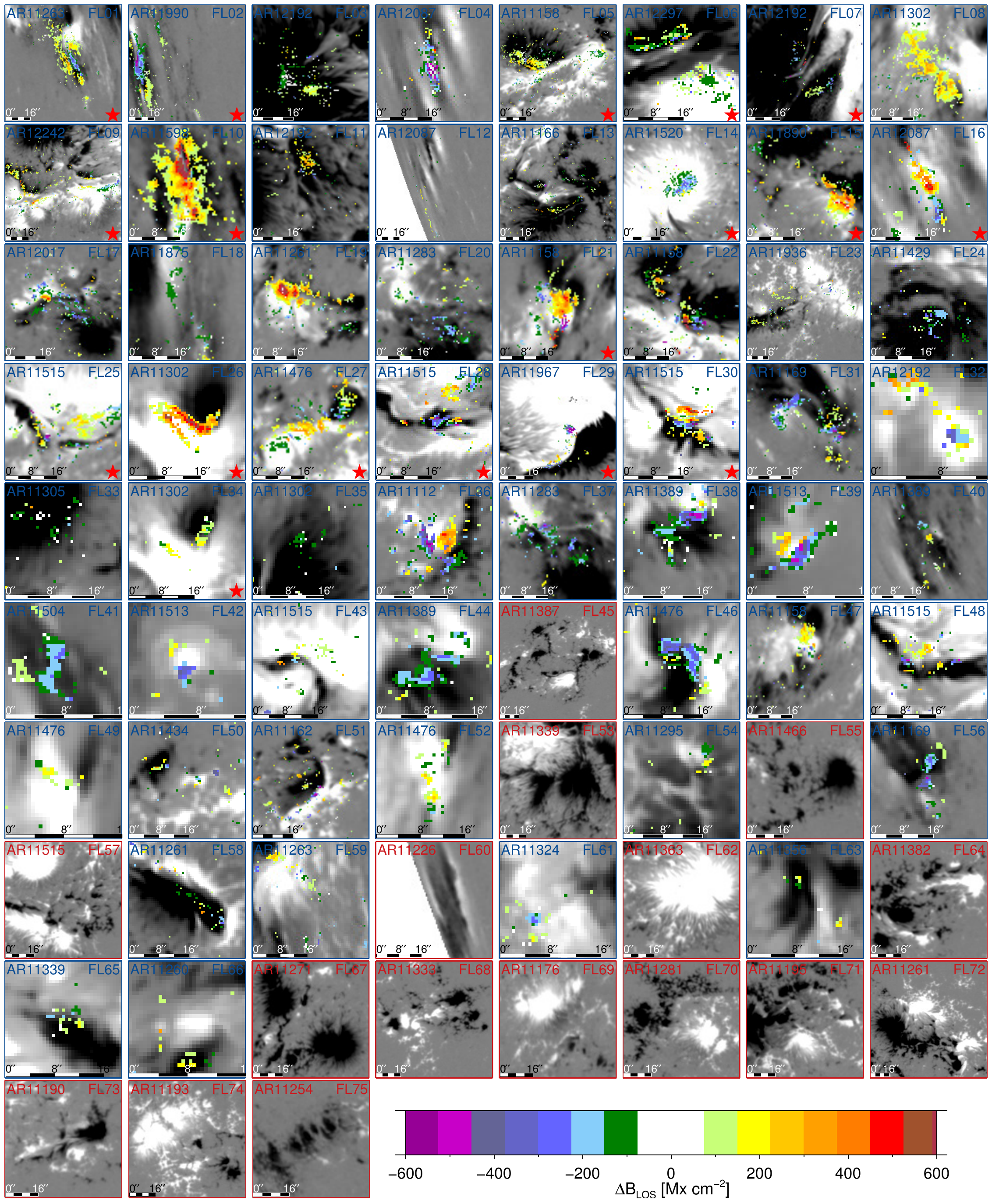} 
\caption{Entire sample of \blos{} 
{All flare-related changes of the evolution of \bt{}} are plotted on magnetograms. Color-coded pixels are locations with \blos{} clipped at $ \pm $600 \Mx{} according to the color bar on the bottom. The magnetogram image in the background was taken at the peak of each flare, and clipped at $\pm$800 \Mx{}. Blue axes and labels are used for flares where \blos{} were found. Red axes and labels represent flares without \blos{}. Image scales are shown by the black and white dashed bars with a length of 16\arcsec{} divided into four parts of 4\arcsec\ each. The FoVs correspond to the black boxes in figure \protect\ref{fig:int_kernelchanges_full}. {Red stars denote the 18 flares that in addition to classical stepwise changes, also showed other types of temporal evolutions of \bt{} (see section \ref{sec:type})}. It is visible that \blos{} occur in kernels and preferentially near opposite polarities. \label{fig:mag_kernelchanges_zoom}}
\end{figure*}

\section{Results}\label{sec:result}
In the following sections, we show properties of \blos{}. We analyze their relation to the strength of the flare (\S\ref{sec31}),  the location (\S\ref{sec32}), the variation of the morphology of the AR (\S\ref{sec33}), their area (\S\ref{sec34}), their distribution with respect to the PIL (\S\ref{sec:pil}), their magnetic flux (\S\ref{sec36}), their relation to {\it GOES} times (\S\ref{sec37}), their rate of change (\S\ref{sec38}), and their frequency distribution (\S\ref{sec39}).

\subsection{Dependence on Flare Strength}\label{sec31}
We found permanent changes in all 18 X-class flares and 92\% (35/38) of the M-class flares. The three M-class events without \blos{} were weak with classes $\le$ M1.5. We observed \blos{} in 35\% (6/17) of the C-class flares. The only B-class event in our sample did not show evidence of \blos{}. We conclude that energetic flares are more likely to show permanent changes.

\subsection{Location Within the Active Region}\label{sec32}
We created masks of different features in the AR to classify the locations of \blos{}, for example, to investigate whether they predominantly occur in umbrae or penumbrae. We removed the limb darkening from the continuum images and calculated the quiet Sun continuum intensity at disk center (I$_{\rm QS}$). The umbra was defined as all pixels with I$_{\rm Umbra} \leq$ 0.6 I$_{\rm QS}$. Penumbral areas corresponded to pixels with intensities between 0.6 I$_{\rm QS}<$ I$_{\rm Penumbra}< 0.92$ I$_{\rm QS}$, {plus pixels added based on a criterion on the magnetic field strength. } To include pixels at the edges of penumbrae with similar intensities as intergranular lanes, which cannot be distinguished based only on continuum intensity, we added pixels with magnetic {flux} densities $|$B$_{\rm LOS}| \geq$ 600 \Mx{} that were not already part of the umbra. This condition for the penumbra on average added $2.3\%\pm2.1\%$ of the total area of the penumbra for 47 different ARs. 

We created three other masks to trace the location of magnetic structures with intensities similar to I$_{\rm QS}$: areas with medium magnetic field strength were defined as 200 \Mx{} $\leq |$B$_{\rm LOS}| <$ 600 \Mx{}, and areas with weak magnetic fields as 30 \Mx{} $\leq |$B$_{\rm LOS}| <$ 200 \Mx{}.  Places with $|$B$_{\rm LOS}| < 30$ \Mx{} were assigned to the quiet Sun. 

Using these masks, we determined that 18.9\% of \blos{} were located in the umbra, 34.3\% in the penumbra, 18.2\% in medium field strengths, 24.2\% in weak field strengths, and 4.3\% in the quiet Sun. \blos{} were mostly observed near the PIL, and regions near the edges between umbra-penumbra and penumbra-medium field strengths. Figure \ref{fig:mag_kernelchanges_zoom} shows a zoom of color-coded \blos{} clipped at $\pm$600 \Mx{} on magnetograms clipped at $\pm$800 \Mx{}. All flares with permanent changes showed that \blos{} were usually located in compact regions or kernels.

\subsection{Changes of the Morphology of the AR}\label{sec33}
Previous studies found changes in the penumbra of ARs that were attributed to flares \citep{Kosovichev2001, Wang2002, Wang2004a, Deng2005, Liu2005,Chen2007}. Twenty four of our 75 analyzed events showed visible changes in the structure of ARs related to flares. 
Penumbral regions disappeared during 15 flares (7 X-, 6 M-, and 2 C-class),  while new penumbral regions appeared after eight flares (2 X-, 4 M-, and 2 C-class). Motions of the PIL were observed during one X- and one M-class event.  

\subsection{Area of \blos{}}\label{sec34}

{In this section, we investigate the size of the areas where the evolution of \bt{} changed at the time of the flare. For this reason, we kept all classes shown in figure \ref{fig:classes}, i.e. all 27,153 changes. We convert the HMI pixel sizes (0\farcs504$\times$0\farcs504) to absolute units (Mm) by taking into account the reference solar radius used by HMI (6.96$\times10^8$ m), the plate scale of HMI continuum images, and the observed solar radius in arcsecs reported by {\it SDO} for the first image in the time range.}
The \blos{} area corrected for foreshortening (${A}_{\BLOS{}}= A/\mu$ where $A$ is the measured area) ranges from {3.4 Mm$^{2}$ to 589.2 Mm$^{2}$} (see Table \ref{tab_sample}, column 8). 

\begin{figure}[tb]
\begin{center}
\includegraphics[width=.49\textwidth]{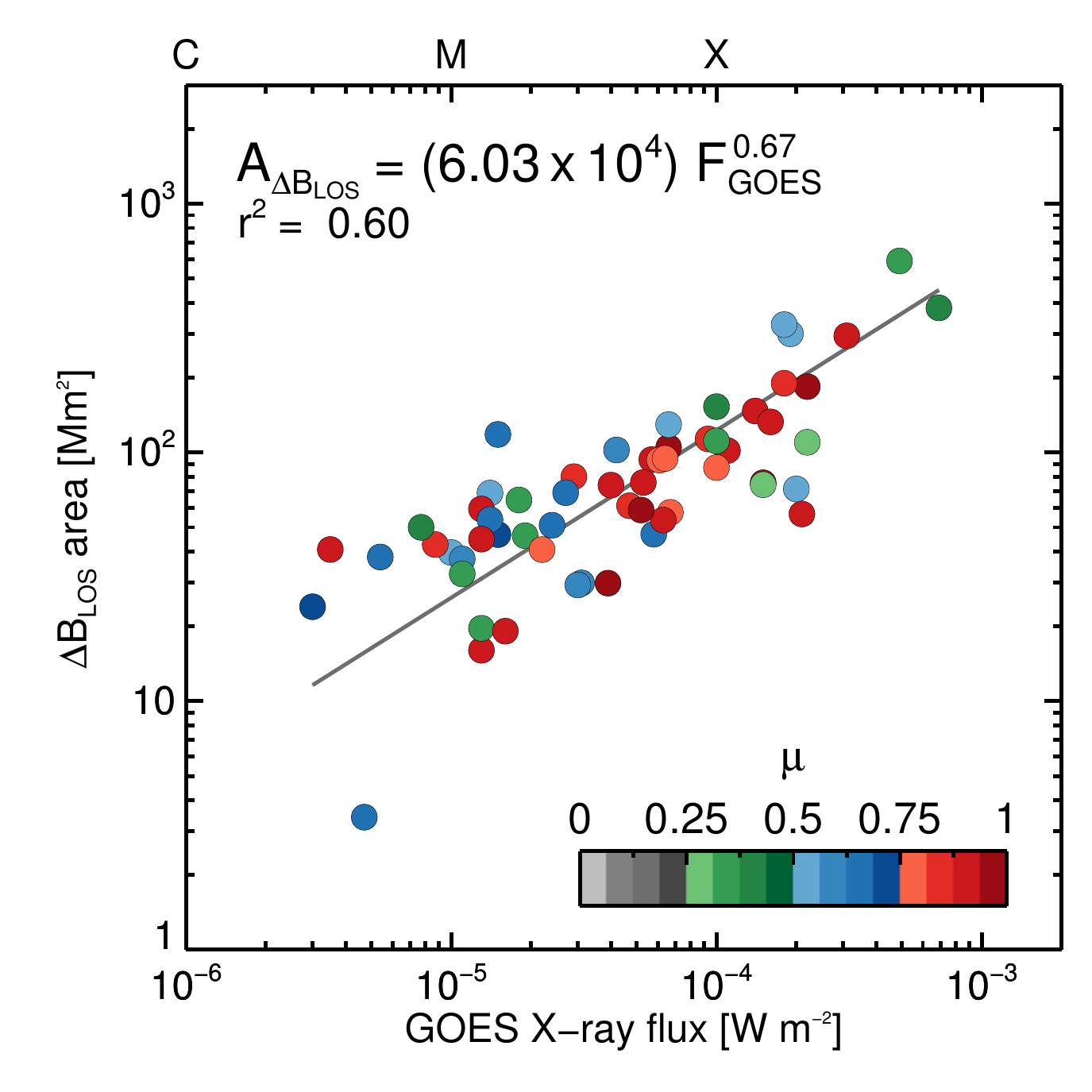} 
\caption{Area corrected for foreshortening of \blos{} as a function of the peak {\it GOES} X-ray flux. Color-coded circles denote the center-to-limb distance $\mu$ of each event. The line is the best fit to a power law with a correlation coefficient of $r^2=0.60$. \label{fig:areachanges}}
\end{center}
\end{figure}

\begin{figure*}[tb]
\begin{center}
\includegraphics[width=.95\textwidth]{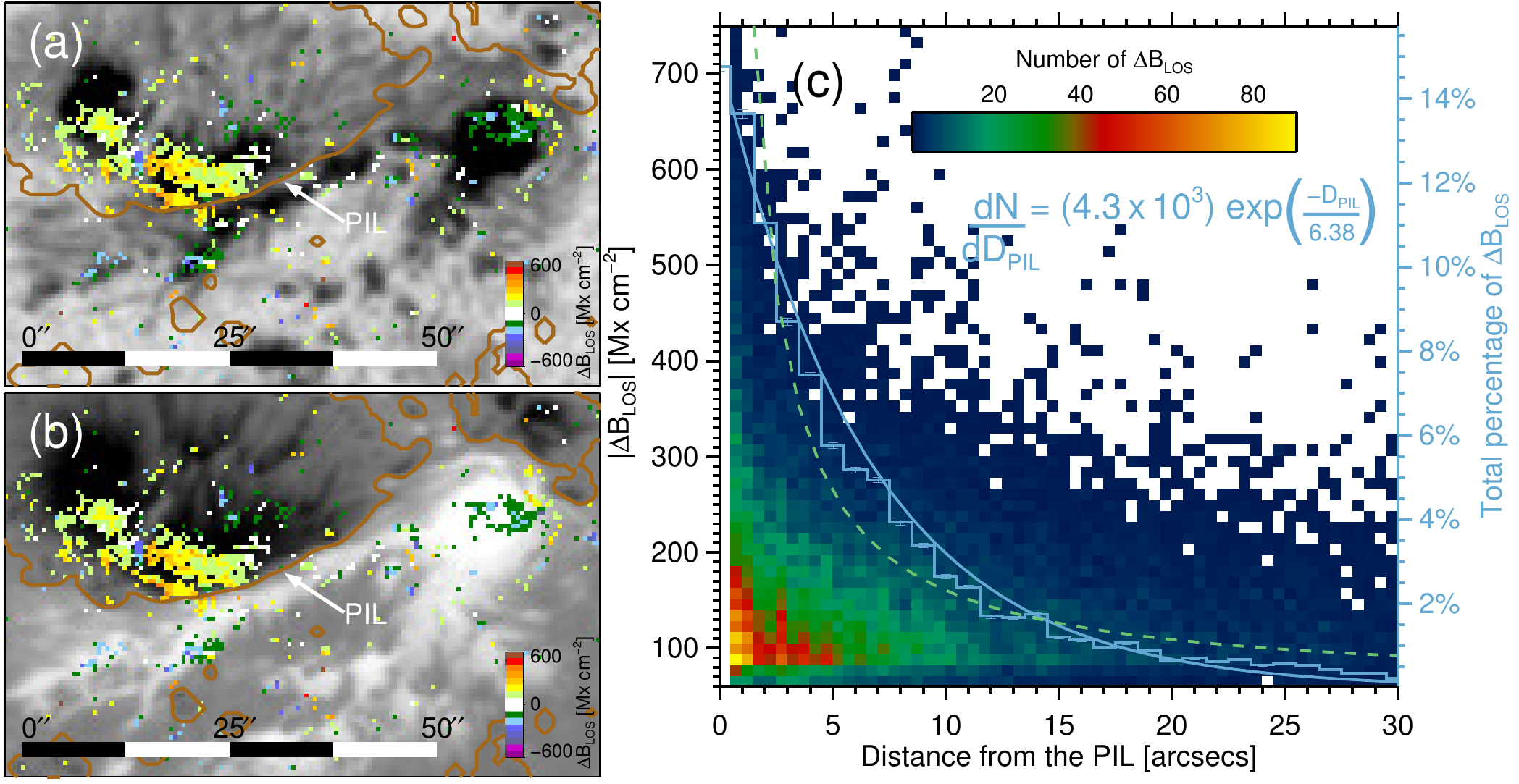}
\caption{Relation between the PIL and \blos{}. {\it Left}: location of \blos{} (color-coded pixels) and the PIL (brown line) during the X2.2 -- \texttt{SOL2011-02-15T01:56} event. The continuum intensity image (panel a) and the magnetogram (panel b) during the maximum of the flare are shown in the background. The image scale is shown by the black and white dashed bar with a length of 50\arcsec{} divided into four parts of 12\farcs5. \blos{} are clipped at $ \pm $600 \Mx{}  according to the vertical color bar. {\it Right}: (c) plot showing the number and strength of \blos{} versus distance to the PIL (${D}_{{\rm PIL}}$). The color scale illustrates the number of \blos{} at a specific distance from the PIL. The histogram (light blue) shows the frequency distribution of \blos{} at certain distances from the PIL.  The solid blue and dashed green lines show the best fit of an exponential and power-law model, respectively. A Kolmogorov--Smirnov test, which can be used as a goodness-of-fit test, showed that the frequency distribution $ d{N}/d{D}_{{\rm PIL}}$ more likely follows an exponential decay given by the equation in the plot.  \label{fig:distance2pil}}
\end{center}
\end{figure*}

Figure \ref{fig:areachanges} shows $ {A}_{\BLOS{}} $ as a function of the {peak {\it GOES} X-ray flux} (${F}_{{GOES}}$). The circles are color-coded with the center-to-limb location of the flares. {A power-law model is given by $f(F)=\kappa F^{\delta}$, where $\kappa$ is a proportional constant, and $\delta$ is the exponent. In the log-log plot the} straight line {is} the best-fit of a power- aw showing that the class of the flare is correlated with the perturbed areas. \blos{} areas in units of {Mm$^2$} are related with the {\it GOES} soft X-ray (SXR) flux from 1 to 8 \AA{} following {a} power-law given by 
\begin{equation}\label{eq:areablos}
{{A}_{\BLOS{}} = (6.03\times10^{4})\ {F}^{0.67}_{{GOES}}},\\ 
\end{equation}
where the power-law exponent is ${\delta_{{A}({\BLOS{}})}=0.67}$ with a correlation coefficient of $r^2=0.60$. We conclude that the \blos{} area is larger for more energetic flares, and is correlated with the peak {\it GOES} X-ray flux.

\subsection{The Relation of \blos{} with the PIL}\label{sec:pil}

The PIL is defined as the region that separates positive and negative magnetic field polarities. It may be related to the flare production, especially when strong fields and high magnetic field gradients are nearby \citep[e.g.,][]{Schrijver2007}. The location of \blos{} was observed to be related to the PIL \citep[e.g.,][]{Wang1994, Kosovichev2001, Wang2010, Wang2012, Petrie2012, Petrie2013, Sun2012, Burtseva2015, Kleint2017}. {In this and the following sections, we restricted our sample by omitting the non-stepwise changes (5.5\% of the sample; types f,g,h, and k).} To detect major PILs in ARs at the maximum of the flare, we traced all pixels at the boundary between opposite magnetic polarities, {and defined a circle with radius 5\arcsec\ centered at each pixel. We defined a pixel} to be part of the PIL if it {is  satisfied} that pixels of each polarity with $|{B}_{\rm LOS}| > 30$ \Mx{} cover at least 10\% of the area of the circle. We selected this criterion to focus on locations where strong polarities are close to each other, and we purposely avoid weak mixed fields. We then determined the closest distance of \blos{} to the nearest PIL. The panels on the left in figure \ref{fig:distance2pil} illustrate the location of the \blos{} during an X2.2 flare as color-coded pixels and the PIL as the brown line. 

In our whole sample of \blos{}, 70\%, 80\%, and 90\% of \blos{} were located no further than 8\farcs2, 11\farcs4, and 18\farcs5 from the PIL, respectively. 3.9\% of \blos{} are located on the PIL at the {\it GOES} peak of the flares. Larger \blos{} tended to be even nearer to the PIL. In fact, 70\%, 80\%, and 90\% of \blos{} $ > $ 250 \Mx{} were located no further than 3\farcs7, 5\farcs5, and 9\farcs4 from the PIL, respectively. The right panel in figure \ref{fig:distance2pil} shows the number and strength of \blos{} as a function of the distance from the PIL (${D}_{{\rm PIL}}$). The color scale represents the number of occurrences of \blos{} ranging from 1 (dark blue) to 90 (light yellow). The highest occurrence, with more than 50\% of all \blos{} is located no further than 5\arcsec{} and with typical absolute values of \blos{} smaller than 200 \Mx{}.  The blue histogram with its axis on the right in the same panel shows the frequency distribution of \blos{} at a distance \dpil{} from the PIL ($ d{N}/d{D}_{{\rm PIL}}$), with bin sizes of 0\farcs5. The frequency distribution of \blos{} from the PIL decays following an exponential behavior, $f(x)=\xi\exp({-}{x/\nu})$, where $\xi$ is a proportional constant, and $\nu$ is the decay exponent. The observational relation is given by 
\begin{equation}\label{eq:expodistpil}
{\frac{dN}{d{D}_{{\rm PIL}}} =(4.3\times10^{3}) \exp\left (\frac{-{D}_{{\rm PIL}}}{6.38}\right ),}
\end{equation}
where the decay exponent is $\nu_{{\rm D}}=6$\farcs{38}. The meaning of $\nu_{{\rm D}}$ is that the number of \blos{} decreases to $ 1/e $ times its initial value at a distance of 6\farcs{38} from the PIL.  $\nu_{{\rm D}}$ represents the mean distance of \blos{} from the PIL. 
{\blos{} reach farther from the PIL in flares~$\geq$M5 (median distance equal to 4\farcs7) than in flares~$<$M5 (median distance equal to 3\farcs6).} 
\subsection{Magnetic Flux Change}\label{sec36}

\citet{Petrie2010} found that the positive magnetic field changes were two times larger than negative changes, and the unsigned magnetic flux tended to decrease after flares. 
The integrated signed change of magnetic flux for each flare  is positive if the absolute value of the field strength increases ($\Phi_{\BLOS{}}^+$), and negative if it decreases ($\Phi_{\BLOS{}}^-$), i.e.,
\begin{equation}
\begin{aligned}
\Phi_{\Delta{B}_{{\rm LOS}}}^+ &= \sum_i \Delta{B}_{{\rm LOS}}(i)*[{A}/\mu(i)],\ \forall \Delta{B}_{{\rm LOS}}>0,\\
\Phi_{\Delta{B}_{{\rm LOS}}}^- &= \sum_i \Delta{B}_{{\rm LOS}}(i)*[{A}/\mu(i)] ,\ \forall \Delta{B}_{{\rm LOS}}<0.
\end{aligned}
\end{equation}

We transformed from flux density (\Mx{}) to flux (Mx) by multiplying the value of the magnetic field times the area corrected for foreshortening ($A/\mu$).
The integrated unsigned flux change for each flare is the sum over all pixels $i$ of the absolute value of positive and negative changes, and is given by
\begin{equation}\label{eq5}
\Sigma|\Phi_{\Delta{B}_{{\rm LOS}}}^{\pm}| = \sum_i |\Delta{B}_{{\rm LOS}}(i)|*[{A}/\mu(i)].
\end{equation}

\begin{figure}[hbt!]
\includegraphics[width=.48\textwidth]{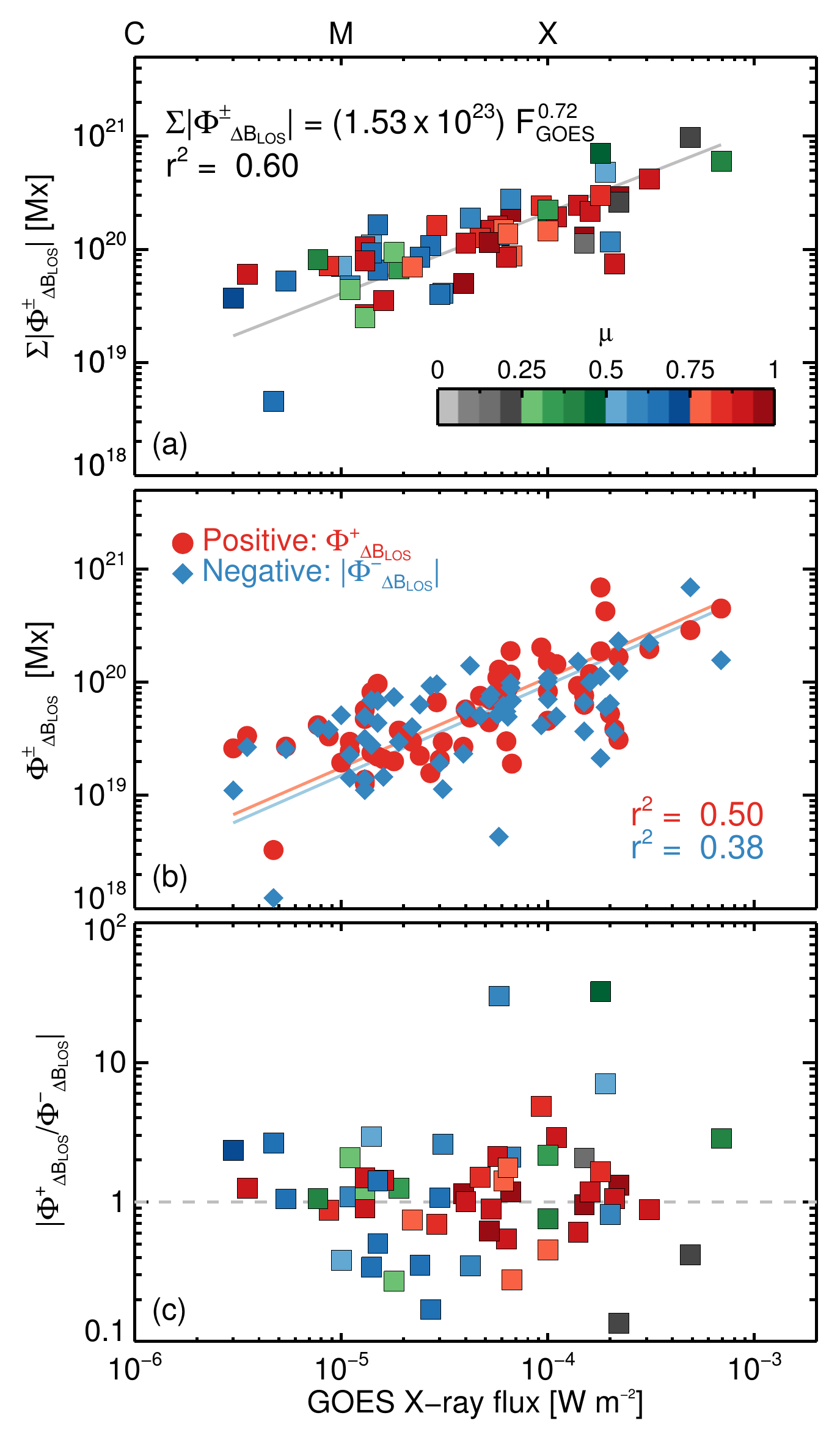}
\caption{Integrated magnetic flux change as a function of the peak {\it GOES} X-ray flux of each flare. {\it Top}: integrated unsigned magnetic flux change $\Sigma|\Phi_{\Delta{B}_{{\rm LOS}}}^{\pm}|$ as a function of the peak {\it GOES} X-ray flux. The color-coded squares denote the center-to-limb distance $\mu$ of each event. {\it Middle}:  integrated signed magnetic flux change $\Phi_{\Delta{B}_{{\rm LOS}}}^{\pm}$ as a function of the peak {\it GOES} X-ray flux. Red circles and blue diamonds illustrate the total positive and negative flux change, respectively. Straight lines are the best-fit of a power-law model $\Phi({F_{{GOES}}})\propto {F}_{{GOES}}^{\delta}$, with exponent ${\delta_{|\Phi|}=0.72}$. {\it Bottom}: ratio between the integrated positive and negative changes. A logarithmic scale is used to better illustrate the ratios between 0 and 1. Squares are color-coded identically to panel (a). More energetic flares tend to change larger magnetic fluxes, but the scatter in the ratio  $|\Phi^{+}_{\Delta{B}_{{\rm LOS}}}/\Phi^{-}_{\Delta{B}_{{\rm LOS}}}|$ is large, which indicates that positive or negative changes do not depend on the flare class.  \label{fig:totalchang2goes2signed}}
\end{figure}

$\Sigma|\Phi_{\BLOS{}}^{\pm}|$ ranges from $4.5\times10^{18} $ Mx to $9.7 \times10^{20} $ Mx (see Table \ref{tab_sample}, column 9).
Figure \ref{fig:totalchang2goes2signed} shows the integrated unsigned ($ \Sigma|\Phi_{\BLOS{}}^{\pm}| $, top) and signed ($\Phi_{\BLOS{}}^{\pm}$, middle) magnetic flux change as a function of the peak {\it GOES} X-ray flux of the flares. The colors of the squares in the top panel of figure  \ref{fig:totalchang2goes2signed} illustrate the center-to-limb distance of the flare. In the middle panel, red circles and blue diamonds stand for integrated positive and negative magnetic flux changes, respectively. The bottom panel in figure \ref{fig:totalchang2goes2signed} shows the ratio between the integrated positive and negative change of magnetic flux. $\Sigma|\Phi_{\BLOS{}}^{\pm}|$ {is} related to the peak {\it GOES} X-ray flux of the flares following a power-law distribution. The lines in panels (a) and (b) in figure \ref{fig:totalchang2goes2signed} show the best fit to power laws. The observational relation between the integrated unsigned magnetic flux change as function of the peak of the SXR flux is given by
\begin{equation}\label{eq:flux}
{\Sigma|\Phi_{\Delta{B}_{{\rm LOS}}}^{\pm}| = (1.53\times 10^{23})\ {F}^{0.72}_{{GOES}}},
\end{equation}
where the exponent is ${\delta_{|\Phi|}=0.72}$ with a correlation coefficient of $r^2=0.60$. The integrated signed positive and negative flux changes also follow a power-law trend with a larger scatter than the integrated unsigned flux change. We did not find evidence that the positive changes were larger than the negative changes in disagreement  with \citet{Petrie2010} (see figure~\ref{fig:totalchang2goes2signed}c). They also reported a dependence of the \blos{} on their location on the solar disk, with limb flares producing more \blos{} than near-disk-center flares. In our flare sample, we do not find any center-to-limb dependence neither with the number of \blos{}, nor the integrated magnetic flux change  (colors of the squares in the top panel of figure  \ref{fig:totalchang2goes2signed}). {This is discussed in more detail in Sect.~\ref{sec45}}.

 \begin{figure}[!tbh]
 \begin{center}
 \includegraphics[width=.49\textwidth]{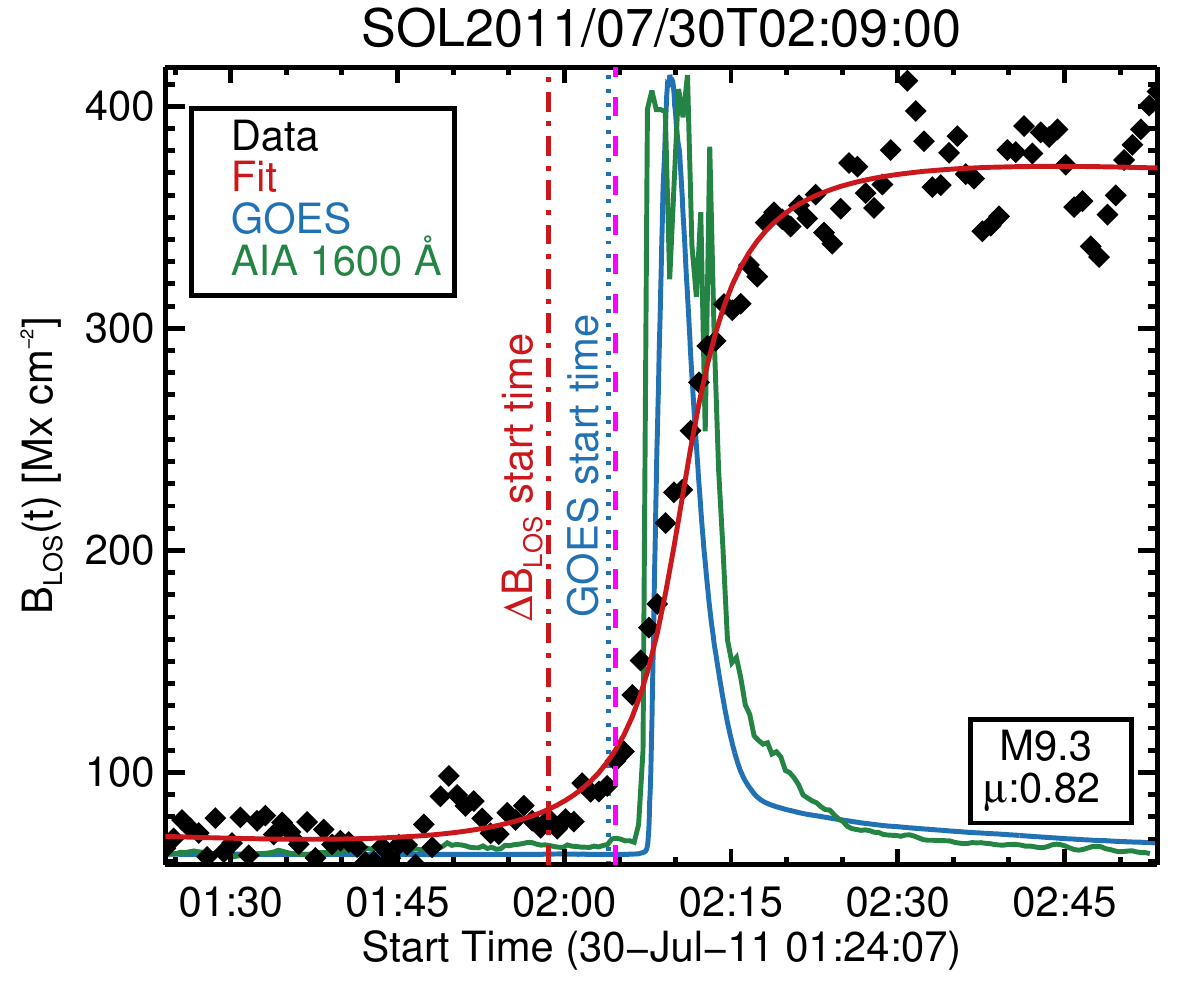}
 \caption{ {{Temporal} evolution of the magnetic field during an M9.3 flare located at $\mu=0.82$. Red, blue, and green lines show the best fit of equation \ref{eq:sudol}, the {\it GOES} 1-8 \AA{} SXR light curve, and the AIA 1600 \AA{} light curve at the same position where the \blos{} took place, respectively. Vertical blue, red, and magenta lines are the {\it GOES} start time, and the \blos{} start time calculated as $t_s=t_0-\pi/n$, and the \blos{} start time calculated as $t'_s=t_0-\pi/(2n)$, respectively. The stepwise change {may have} started before the SXR and chromospheric emission.} \label{fig:allobs}}
 \end{center}
 \end{figure}

\subsection{\blos{} and the {\it GOES} Times}\label{sec37}
\blos{} are a common phenomenon of the  impulsive phase of flares \citep{Cliver2012}. We tested if the duration of the impulsive phase of the flares is correlated with the \blos{}. We used the period between the start and peak times of the {\it GOES} SXR flux as a proxy for the duration of the impulsive phase. We did not find any relation of the length of the impulsive phase and \blos{} for the flares in our sample.

As a next step, we investigated the timing of the changes. We find that 2.2\% of the \blos{} started before the {main} {\it GOES} SXR emission. Figure \ref{fig:allobs} shows a stepwise change during an M9.3 flare located at $\mu=0.82$. The evolution of \bt{} (black diamonds), and the fit of equation \ref{eq:sudol} (red line) show that the change of the magnetic field {may have} started before the {\it GOES} SXR 1-8 \AA\ emission (blue line), as well as the chromospheric AIA 1600 \AA\ emission observed at the same location (green line). We calculated the start time for equation \ref{eq:sudol} as $t_s=t_0-\pi/n$ (vertical red line), which differs from previous studies \citep[e.g.][]{Petrie2010,Burtseva2015}, who estimated the start time as $t'_s=t_0-\pi/(2n)$ (vertical magenta line). With their starting time ($t'_s$), the increase in {\it GOES} flux (vertical blue line) appears to start earlier than the change, but the start of the {\it GOES} event is also an arbitrary definition by the {\it GOES} team as ``the first minute, in a sequence of 4 minutes, of steep monotonic increase.''\footnote{http://www.swpc.noaa.gov}
Figure \ref{fig:allobs} shows that the start of \bt{} may be better described by choosing the start time as $t_s$. {The {\it GOES} light curve in figure \ref{fig:allobs} shows a small bump before the flare onset,which may be an explanation for the apparent early timing of the change if this pixel was affected by the pre-main-flare event. Because {\it GOES} light curves are integrated over the solar disk, we cannot  determine the reason for the apparent timing discrepancy. Nevertheless, with these assumptions/limitations}, 2.2\% of the \blos{} onsets occurred before the {main} SXR emission onset.

\startlongtable
\begin{deluxetable*}{rlcccccccc}
%\tabletypesize{\small }
%\tablecolumns{11} 
\tablewidth{0pt}
\tablecaption{List of flares\label{tab_sample}} 
\tablehead{
&\colhead{Observation}              & \colhead{{\it GOES}} & 
\colhead{{\it GOES}}         & \colhead{NOAA}   & \colhead{Latitude} & \colhead{Longitude} & \colhead{${\rm A}_{\BLOS{}}$}& \colhead{$\Sigma|\Phi_{\BLOS{}}^{\pm}|$} & \colhead{Exponent}  \\
\colhead{Index}&\colhead{Date} & \colhead{Peak} & \colhead{Class} & \colhead{AR}   & \colhead{(deg)}  & \colhead{(deg)} & \colhead{(Mm$^2$)} & \colhead{(10$^{18}$ Mx)} & \colhead{(\Mx{})}}
%\decimalcolnumbers}
\startdata
\textbf{\texttt{FL}01} & 2011 Aug 9  &08:05&  X6.9  &11263&  14.9 N  &  69.6 W  &381.5&604.1&56.3\\
\textbf{\texttt{FL}02} & 2014 Feb 25  &00:49&  X4.9  &11990&  12.0 S  &  81.9 E  &589.2&977.3&100.8\\
\textbf{\texttt{FL}03} & 2014 Oct 24  &21:41&  X3.1  &12192&  11.1 S  &  20.5 W  &294.6&418.9&62.7\\
\textbf{\texttt{FL}04} & 2014 Jun 10  &11:42&  X2.2  &12087&  15.0 S  &  79.8 E  &109.9&261.3&149.7\\
\textbf{\texttt{FL}05} & 2011 Feb 15  &01:56&  X2.2  &11158&  20.9 S  &  10.8 W  &184.1&293.3&78.2\\
\textbf{\texttt{FL}06} & 2015 Mar 11  &16:22&  X2.1  &12297&  17.0 S  &  21.0 E  &56.6&74.8&85.8\\
\textbf{\texttt{FL}07} & 2014 Oct 27  &14:47&  X2.0  &12192&  18.8 S  &  57.4 W  &71.4&117.1&133.0\\
\textbf{\texttt{FL}08} & 2011 Sep 24  &09:39&  X1.9  &11302&  12.0 N  &  59.9 E  &300.3&483.5&75.7\\
\textbf{\texttt{FL}09} & 2014 Dec 20  &00:28&  X1.8  &12242&  19.1 S  &  29.0 W  &189.9&300.7&66.8\\
\textbf{\texttt{FL}10} & 2012 Oct 23  &03:17&  X1.8  &11598&  13.3 S  &  58.9 E  &326.8&709.4&135.4\\
\textbf{\texttt{FL}11} & 2014 Oct 22  &14:28&  X1.6  &12192&  14.4 S  &  13.9 E  &132.8&217.9&70.0\\
\textbf{\texttt{FL}12} & 2014 Jun 10  &12:52&  X1.5  &12087&  17.0 S  &  81.9 E  &74.2&112.9&135.7\\
\textbf{\texttt{FL}13} & 2011 Mar 9  &23:23&  X1.5  &11166&  08.0 N  &  11.8 W  &75.5&129.4&83.3\\
\textbf{\texttt{FL}14} & 2012 Jul 12  &16:50&  X1.4  &11520&  14.0 S  &  03.9 W  &146.8&244.9&90.2\\
\textbf{\texttt{FL}15} & 2013 Nov 10  &05:14&  X1.1  &11890&  13.4 S  &  13.9 W  &101.8&194.3&100.1\\
\textbf{\texttt{FL}16} & 2014 Jun 11  &09:06&  X1.0  &12087&  18.0 S  &  64.9 E  &111.3&223.8&152.7\\
\textbf{\texttt{FL}17} & 2014 Mar 29  &17:48&  X1.0  &12017&  10.2 N  &  32.9 W  &87.0&146.3&73.4\\
\textbf{\texttt{FL}18} & 2013 Oct 28  &02:03&  X1.0  &11875&  04.0 N  &  66.0 W  &152.8&191.8&51.6\\
\textbf{\texttt{FL}19} & 2011 Jul 30  &02:09&  M9.3  &11261&  14.0 N  &  34.9 E  &113.4&244.2&137.5\\
\textbf{\texttt{FL}20} & 2011 Sep 8  &15:46&  M6.7  &11283&  14.0 N  &  39.9 W  &57.4&87.6&67.7\\
\textbf{\texttt{FL}21} & 2011 Feb 18  &10:11&  M6.6  &11158&  21.1 S  &  55.0 W  &129.4&277.7&153.5\\
\textbf{\texttt{FL}22} & 2011 Feb 13  &17:38&  M6.6  &11158&  19.9 S  &  03.9 E  &104.7&215.6&125.3\\
\textbf{\texttt{FL}23} & 2013 Dec 31  &21:58&  M6.4  &11936&  15.5 S  &  36.0 W  &94.6&136.8&68.9\\
\textbf{\texttt{FL}24} & 2012 Mar 9  &03:53&  M6.3  &11429&  15.8 N  &  03.4 W  &53.5&85.2&55.8\\
\textbf{\texttt{FL}25} & 2012 Jul 5  &11:44&  M6.1  &11515&  18.7 S  &  32.6 W  &93.2&149.8&100.8\\
\textbf{\texttt{FL}26} & 2011 Sep 24  &20:36&  M5.8  &11302&  13.0 N  &  52.0 E  &46.9&133.4&180.8\\
\textbf{\texttt{FL}27} & 2012 May 10  &04:18&  M5.7  &11476&  12.9 N  &  22.0 E  &93.7&161&78.5\\
\textbf{\texttt{FL}28} & 2012 Jul 4  &09:55&  M5.3  &11515&  17.3 S  &  18.0 W  &75.8&147.8&111.8\\
\textbf{\texttt{FL}29} & 2014 Feb 4  &04:00&  M5.2  &11967&  12.9 S  &  06.9 W  &58.7&115.4&125.7\\
\textbf{\texttt{FL}30} & 2012 Jul 5  &03:36&  M4.7  &11515&  18.3 S  &  29.2 W  &61.1&126.2&133.4\\
\textbf{\texttt{FL}31} & 2011 Mar 14  &19:52&  M4.2  &11169&  16.1 N  &  49.6 W  &102.3&188.8&106.5\\
\textbf{\texttt{FL}32} & 2014 Oct 24  &07:48&  M4.0  &12192&  20.7 S  &  07.3 W  &73.9&114.2&73.1\\
\textbf{\texttt{FL}33} & 2011 Oct 2  &00:50&  M3.9  &11305&  09.0 N  &  12.0 W  &29.8&50.2&85.8\\
\textbf{\texttt{FL}34} & 2011 Sep 24  &17:25&  M3.1  &11302&  13.0 N  &  54.0 E  &29.9&40.9&47.1\\
\textbf{\texttt{FL}35} & 2011 Sep 24  &19:21&  M3.0  &11302&  12.7 N  &  42.7 E  &29.3&40&60.5\\
\textbf{\texttt{FL}36} & 2010 Oct 16  &19:12&  M2.9  &11112&  20.8 S  &  27.9 W  &79.9&162.5&109.7\\
\textbf{\texttt{FL}37} & 2011 Sep 9  &06:09&  M2.7  &11283&  16.0 N  &  47.0 W  &68.8&108.3&79\\
\textbf{\texttt{FL}38} & 2011 Dec 31  &13:15&  M2.4  &11389&  25.9 S  &  46.0 E  &51.0&85.8&93.1\\
\textbf{\texttt{FL}39} & 2012 Jun 29  &09:20&  M2.2  &11513&  17.0 N  &  36.9 E  &40.7&69.9&98.7\\
\textbf{\texttt{FL}40} & 2011 Dec 29  &13:50&  M1.9  &11389&  25.2 S  &  70.7 E  &46.3&66.9&49.5\\
\textbf{\texttt{FL}41} & 2012 Jun 9  &16:53&  M1.8  &11504&  17.0 S  &  73.9 E  &64.4&94&70.9\\
\textbf{\texttt{FL}42} & 2012 Jun 30  &18:32&  M1.6  &11513&  13.9 N  &  18.0 E  &19.1&35.5&92.5\\
\textbf{\texttt{FL}43} & 2012 Jul 6  &08:23&  M1.5  &11515&  17.0 S  &  39.9 W  &118.2&165.1&60\\
\textbf{\texttt{FL}44} & 2011 Dec 31  &16:26&  M1.5  &11389&  25.7 S  &  42.7 E  &46.7&65.7&50.4\\
\textbf{\texttt{FL}45} & 2011 Dec 26  &02:27&  M1.5  &11387&  21.0 S  &  32.9 W  & \nodata & \nodata &  \nodata\\
\textbf{\texttt{FL}46} & 2012 May 8  &13:08&  M1.4  &11476&  12.9 N  &  44.0 E  &53.7&93.3&89.9\\
\textbf{\texttt{FL}47} & 2011 Feb 18  &13:03&  M1.4  &11158&  21.1 S  &  55.8 W  &68.6&109.5&76.2\\
\textbf{\texttt{FL}48} & 2012 Jul 4  &14:40&  M1.3  &11515&  18.0 S  &  18.0 W  &44.8&79.1&91.3\\
\textbf{\texttt{FL}49} & 2012 May 5  &23:01&  M1.3  &11476&  08.9 N  &  74.7 E  &19.6&24.9&74.9\\
\textbf{\texttt{FL}50} & 2012 Mar 17  &20:39&  M1.3  &11434&  20.0 S  &  25.0 W  &16.0&26.6&94.2\\
\textbf{\texttt{FL}51} & 2011 Feb 18  &21:04&  M1.3  &11162&  18.1 N  &  03.7 W  &59.3&106.3&110.3\\
\textbf{\texttt{FL}52} & 2012 May 6  &01:18&  M1.1  &11476&  11.0 N  &  72.9 E  &32.5&44&72.8\\
\textbf{\texttt{FL}53} & 2011 Nov 5  &11:21&  M1.1  &11339&  19.7 N  &  41.1 E  & \nodata & \nodata &  \nodata\\
\textbf{\texttt{FL}54} & 2011 Sep 22  &10:00&  M1.1  &11295&  25.1 N  &  57.2 W  &37.3&47.4&63.4\\
\textbf{\texttt{FL}55} & 2012 Apr 27  &08:24&  M1.0  &11466&  13.0 N  &  29.5 W  & \nodata & \nodata &  \nodata\\
\textbf{\texttt{FL}56} & 2011 Mar 15  &00:22&  M1.0  &11169&  11.8 N  &  83.3 W  &39.6&70.6&98.1\\
\textbf{\texttt{FL}57} & 2012 Jul 3  &03:42&  C9.9  &11515&  18.2 S  &  02.4 W  & \nodata & \nodata &  \nodata\\
\textbf{\texttt{FL}58} & 2011 Aug 3  &07:58&  C8.7  &11261&  15.8 N  &  29.3 W  &42.6&71.2&97\\
\textbf{\texttt{FL}59} & 2011 Aug 8  &22:09&  C7.7  &11263&  17.4 N  &  66.5 W  &50.0&81.2&86.6\\
\textbf{\texttt{FL}60} & 2011 May 27  &16:43&  C5.6  &11226&  21.4 S  &  89.8 E  & \nodata & \nodata &  \nodata\\
\textbf{\texttt{FL}61} & 2011 Oct 20  &15:39&  C5.4  &11324&  11.5 N  &  47.8 E  &38.0&52.6&45.8\\
\textbf{\texttt{FL}62} & 2011 Dec 5  &15:18&  C4.9  &11363&  20.3 S  &  02.4 W  & \nodata & \nodata &  \nodata\\
\textbf{\texttt{FL}63} & 2011 Nov 22  &04:04&  C4.7  &11356&  12.7 N  &  50.2 E  &3.4&4.5&67.3\\
\textbf{\texttt{FL}64} & 2011 Dec 21  &04:55&  C4.3  &11382&  19.4 S  &  02.1 E  & \nodata & \nodata &  \nodata\\
\textbf{\texttt{FL}65} & 2011 Nov 7  &03:10&  C3.5  &11339&  20.3 N  &  22.0 E  &40.7&60.2&59.7\\
\textbf{\texttt{FL}66} & 2011 Jul 27  &10:02&  C3.0  &11260&  18.6 N  &  40.6 E  &24.0&37.1&70.4\\
\textbf{\texttt{FL}67} & 2011 Aug 20  &22:58&  C2.9  &11271&  13.6 N  &  12.0 E  & \nodata & \nodata &  \nodata\\
\textbf{\texttt{FL}68} & 2011 Oct 27  &18:44&  C2.6  &11333&  15.0 N  &  13.5 E  & \nodata & \nodata &  \nodata\\
\tablenotemark{a}\textbf{\texttt{FL}69} & 2011 Mar 31  &15:35&  C2.6  &11176&  16.0 S  &  39.8 W  & \nodata & \nodata &  \nodata\\
\textbf{\texttt{FL}70} & 2011 Sep 3  &07:56&  C2.4  &11281&  21.0 S  &  01.1 W  & \nodata & \nodata &  \nodata\\
\textbf{\texttt{FL}71} & 2011 Apr 23  &07:34&  C2.4  &11195&  16.0 S  &  23.8 E  & \nodata & \nodata &  \nodata\\
\textbf{\texttt{FL}72} & 2011 Aug 1  &12:40&  C2.0  &11261&  18.2 N  &  01.8 W  & \nodata & \nodata &  \nodata\\
\textbf{\texttt{FL}73} & 2011 Apr 12  &03:46&  C1.7  &11190&  11.7 N  &  25.0 E  & \nodata & \nodata &  \nodata\\
\textbf{\texttt{FL}74} & 2011 Apr 18  &19:02&  C1.5  &11193&  16.4 N  &  11.7 E  & \nodata & \nodata &  \nodata\\
\textbf{\texttt{FL}75} & 2011 Jul 16  &17:05&  B6.2  &11254&  24.3 S  &  42.8 E  & \nodata & \nodata &  \nodata
\enddata
\tablenotetext{a}{$\textbf{\texttt{FL}69}$ was excluded from the analysis because of gaps in the data.}
\tablecomments{The details of the flares are shown in columns 1 - 7. {The areas of \blos{} corrected for foreshortening} (A/$\mu$), magnetic flux changes of \blos{}, and exponent of the \blos{} frequency distribution are in columns 8 - 10 if the flare showed permanent changes.}
\end{deluxetable*}

\begin{figure*}[tb]
\begin{center}
\includegraphics[width=1.\textwidth]{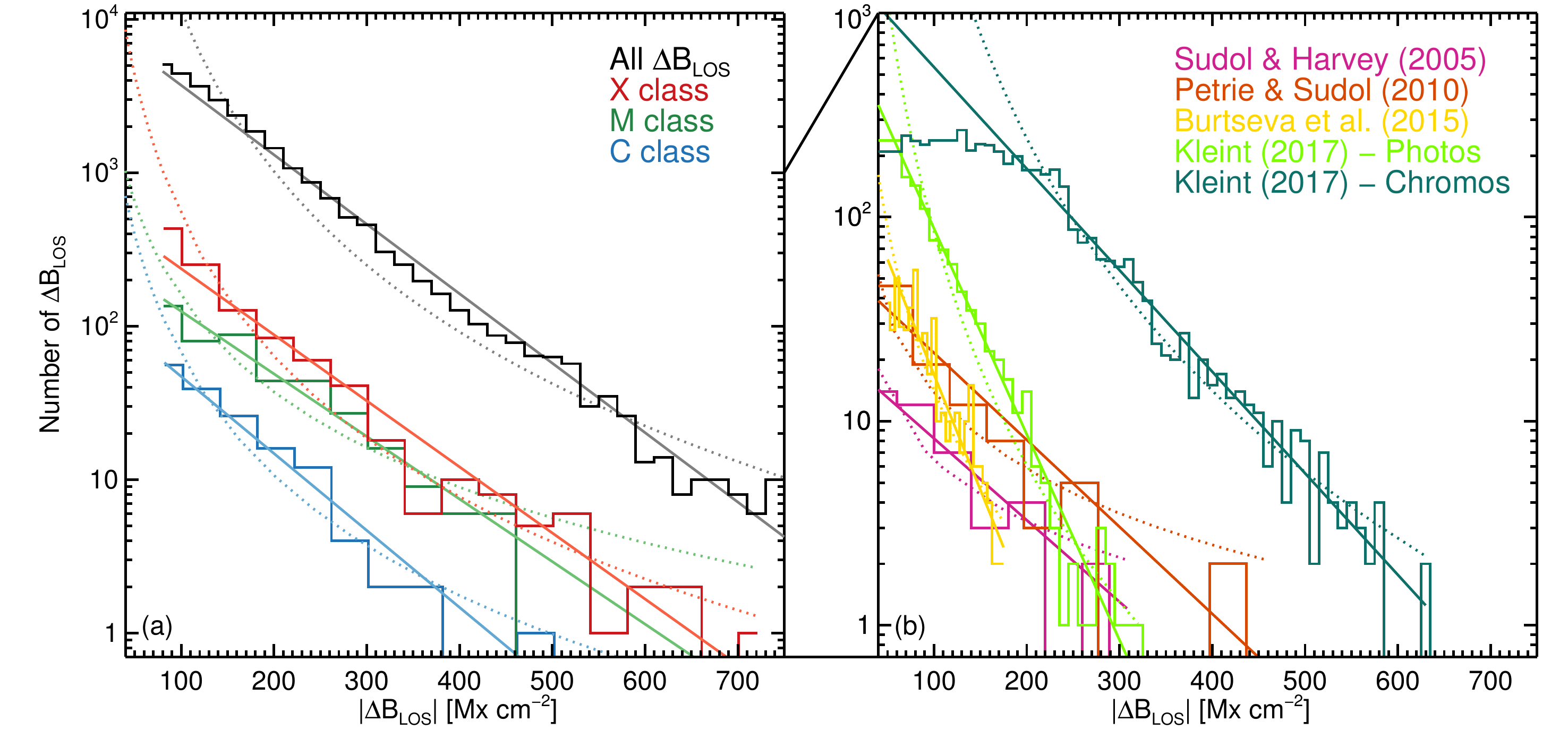}
\caption{{\it Left}: histograms of the frequency distribution of the sizes of \blos{} during three example flares: C7.7 -- \texttt{SOL2011-08-08T22:09} (blue), M4.2 -- \texttt{SOL2011-03-14T19:52} (green), X4.9 -- \texttt{SOL2014-02-25T00:49} (red), and the whole sample of \blos{} (black). {\it Right}: the reconstructed histograms of the \blos{} frequency distribution from \protect \citet{Sudol2005} ({purple}), \protect \citet{Petrie2010} ({orange}), the X8.3 flare studied by \protect\citet{Burtseva2015} ({yellow}), and \protect \citet{Kleint2017} ({light green}: photospheric \blos{}; and {dark green}: chromospheric \blos{}). We binned  \citet{Petrie2010}'s data to 40 \Mx{} as in \citet{Sudol2005}. The distributions of the events of \protect \citet{Burtseva2015} were not rebinned, due to their low dynamic range. {The binning for the left panel and for \citet{Kleint2017} is arbitrary for display purposes.}
The solid and dotted lines show the best fit of an exponential and power-law models, respectively. {The sizes of \blos{} from this study, as well as from previous studies, follow a decreasing exponential distribution. } 
\label{fig:statistics}}
\end{center}
\end{figure*}

\subsection{Rate of Changes of \blos{}}\label{sec38}
How fast does \bt{} change? We estimated the rate of magnetic changes ($\BLOS{/\Delta t}$) taking the duration of the permanent change as $\Delta t=\pi n^{-1}$ according to \citet{Sudol2005} {and not our duration above, to make our results comparable}. The shortest duration of \blos{} in our study is the  45 s HMI cadence, though it does not mean that the changes may not occur in shorter times. Therefore, we set the lower limit for the duration of the change from the fit of stepwise function (equation \ref{eq:sudol}) to 45 s. Less than 10\% of \blos{} were not well fitted with equation 1 mainly because {of} the {different} types of permanent changes (see \S\ref{sec:type});  therefore we adopted the interval between $\pm5$ minutes before and after the start and end {\it GOES} times for each flare as the maximum possible value for the duration of the \blos{}. The rate of change spans from 0.2 \Mx{} min$^{-1}$ to 307 \Mx{} min$^{-1}$. The mean of the rate of change is 20.7 \Mx{} min$^{-1}$. 91\% of \blos{} have a rate of change smaller than 50 \Mx{} min$^{-1}$ consistent with previous findings \citep{Sudol2005}. 
 
\subsection{Frequency Distribution of the Sizes of \blos{}} \label{sec39}
Weak permanent changes $\lesssim$ 100 \Mx{} were observed more often than larger \blos{} \citep{Sudol2005, Petrie2010}. Here, we investigate the frequency distribution of strong and weak \blos{} in our sample. The permanent changes in our sample ranged from 80 Mx cm$^{-2}$ (selected lower limit) to $\sim$750 \Mx{}. Previous studies found photospheric \blos{} up to 450 \Mx{} \citep{Petrie2010, Cliver2012, Johnstone2012, Burtseva2015, Kleint2017}. Only 2\% of \blos{} in our sample are larger than 450 \Mx{}. The black histogram in the left panel of figure \ref{fig:statistics} shows the frequency distribution of all permanent changes in our sample ($d{N}/d\BLOS{}$). The straight line is the best fit of an exponential model, $d{N}/d\BLOS{} = A\exp\left({-}|\BLOS{}|/\tau \right) $. {We used a maximum-likelihood fitting of univariate distributions to fit the data \citep{Venables2002}}. The observational relation of the frequency distribution for all permanent changes is given by
\begin{equation}\label{eq:distribution}
{\frac{d{N}}{d\BLOS{}}=(1.0\times10^{4})\ \exp\left (-\frac{|\BLOS{}|}{88.4}\right )}.
\end{equation}

The number of \blos{} reduces to $1/e$  times their initial value when $\BLOS{}={88.4}$ \Mx{}, i.e. the mean \blos{} amplitude {(with our selected lower limit)} is {88.4} \Mx{}. The colored histograms in the left panel in figure \ref{fig:statistics} show the frequency distribution of an X- (red), an M- (green), and C-class (blue) flare. {We found that for each single flare and for the entire sample, weak \blos{} are  observed more often, and the frequency distribution of the strength of \blos{} follows an exponential distribution.  }

\section{Discussion}

In the following sections, we will discuss the results obtained in Sections \ref{sec31} to \ref{sec39}.

\subsection{\blos{} and Strength of the Flares}

We found that all flares with classes larger than M1.6 showed \blos{}, and 11 of 14 flares from M1.5 to M1.0 showed \blos{}. Our result expands the previous statistical study by {\citet{Petrie2010}} who found \blos{} for all flares with classes $\ge$ M5 {(their selected lower limit)}. Six of 17 C-flares showed \blos{}. The smallest flare with \blos{} is the C3.0 class flare \texttt{SOL2011-07-27T10:02}. To our knowledge, this is the smallest flare with \blos{} reported {to date}, replacing the C4.7 flare previously reported by \citet{Wang2012b}.

We observed that only a third of the C-class flares showed \blos{}, which agrees with a previous study that used {\it SDO}/HMI data as well \citep{Wang2012b}. {\it GOES} classes are an arbitrary classification, defined as an order of magnitude in the SXR (1-8 \AA) flux (W m$^{-2}$). Therefore, the lower fraction of C-class flares where changes were observed may be biased because of observational limits but not based on physics, which we will test with the following calculation.

For example, a C1.0 flare will show a total area of permanent changes of {${{\rm A}_{\BLOS{}}\approx5.8\ \rm{Mm}^2\approx}$(11\arcsec)$^2$}, based on equation~\ref{eq:areablos}.  If our theoretical flare were located at $\mu=1$, the area {$\approx5.8\ \rm{Mm}^2$} corresponds to {$N_{\BLOS{}}\approx44$} HMI pixels . For a flare at $\mu=0.5$, only {$\approx22$} pixels would appear changing. From equation \ref{eq:flux}, the total unsigned magnetic flux change would be {$\Sigma|\Phi^{\pm}_{\BLOS{}}|\approx7.6\times10^{18}$} Mx. The magnitudes of the changes will be distributed exponentially according to equation~\ref{eq:distribution}. This implies that smaller changes are more frequent. Using equation~\ref{eq5} and assuming 0\farcs$5=360$ km, this corresponds to an average unsigned magnetic field change per pixel of {$7.6\times10^{18} / ((3.6\times10^{7})^2)/N_{\BLOS{}} = 133$ \Mx{}}. This means that one would be looking for {fewer} than a dozen large permanent changes ($>$150 \Mx{}) in the FoV, and even {fewer} if the AR is located away from disk center. Finding these changes in the on average $\sim1.5\times10^5$ pixels in the FoV may prove difficult. For this reason, we believe that \blos{} may be a common phenomenon also during small flares, and the lack of \blos{} in C-class flares is likely attributed to the spatial resolution of our data, and the selection limit used in this study.

\subsection{\blos{} and their Location in the AR}

Flares have been reported to change the photospheric structure in single events, and we found that changes of the penumbra are relatively common. In 23 of 75 events, the penumbra changed. We observed that \blos{} tend to occur together in compact regions, and most of them (71\%) occurred within parts of the AR with magnetic flux densities larger than 200 \Mx{}.

From the observations of disappearing penumbrae or from visible changes of loop structures \citep[see e.g. movies in][]{suetal2013}, we expect that field lines are tilted during the flare process. To produce a given \blos{}, a strong field will require a smaller tilt angle than a weak field. For example, to observe a \blos{} $=$ 80 \Mx{} from a LOS field strength of 1300 \Mx{}, a tilt of $\cos^{-1}$(1220/1300) = 20$^\circ$ is required. For a weaker field of 200 \Mx{}, a tilt of 53$^\circ$ would be required. While no quantitative conclusion can be given here, smaller tilts seem to be observed more often at footpoints in {\it SDO}/AIA images in our experience, but a thorough study is missing.

About half of all magnetic field changes occurred in sunspots in our sample (34.3\% in the penumbra and 18.9\% in the umbra). Our sample of flares showed that places with morphological changes observed in the continuum images were also associated with \blos{}.  These types of morphological changes were explained as a permanent change in the inclination of the penumbral magnetic field \citep{Liu2005, Chen2007}. However, in this study, we analyzed the LOS component of the field without any information about the orientation of the magnetic field. 

The remaining 50\% of the changes were located in the network and internetwork with LOS field strengths that range from 30 \Mx{} to 600 \Mx{}. 4.3\% were observed in the quiet Sun. We do not have an explanation for their appearance in these locations of relatively small magnetic flux.

\subsection{Area of \blos{}}

We found that the area where permanent changes occur tends to be larger in stronger flares. Because strong flares are also more energetic \citep[e.g.,][]{Wang2009, Aschwanden2015}, it is plausible that more energy is available to be converted into magnetic field changes. 

\subsection{{\blos{} and the PIL}}

We found that the number of changes decays exponentially with distance from the PIL. \blos{} in the neighborhood of PILs have been found in diverse studies \citep{Wang2010, Petrie2012, Sun2012, Burtseva2013, Petrie2013, Petrie2016}. In this study, we gave the first quantification of the distance of \blos{} with respect to the PIL.

We also observed that the permanent changes occur closer to the PIL for more energetic flares, which may be related to the results of \citet{Schrijver2007}, who found that all large flares were associated with strong gradient PILs. \citet{Wang1992, Wang1994} found that the shear angle, which is the angle between the observed magnetic field and the calculated potential field, increased during their analyzed flares. 
The proximity of \blos{} to the PIL could indicate a change in the shear after the flare process.

\subsection{{The Magnetic Flux Change During \blos{}}}\label{sec45}

We found that larger flares show a larger signed and unsigned magnetic flux change. Contrary to \citet{Burtseva2013}, we did not find any difference between the positive and negative flux changes: their ratio is not correlated with the {\it GOES} class of the events. 
We also found that the magnetic flux change does not depend on $\mu$ (figure~\ref{fig:totalchang2goes2signed}), which disagrees with \citet{Petrie2010}, who observed larger changes near the limb and associated them with a change in the inclination of the field. {Our flare sample slightly differs from others; for example, \citet{Burtseva2013} and \citet{Petrie2010} restricted their analysis to flares $\ge$M5 observed by GONG, whose spatial sampling (2\farcs5 pixel$^{-1}$) is different to HMI's (0\farcs5 pixel$^{-1}$) and they omitted flares close to the limb with a longitude $\ge$65$^\circ$ from the central meridian. Additionally, their flare sample was from the previous solar cycle 23, while our sample is based on cycle 24.}
{For a more direct comparison with previous studies, we recalculated figures \ref{fig:areachanges}, \ref{fig:totalchang2goes2signed}, and \ref{fig:statistics} by restricting our flare sample to match their criteria. We omitted the 9.2\% of \blos{} that are not well described by equation \ref{eq:sudol} (2498 of 27,153 \blos{}), and the 10 flares close to the limb (5 X-, 4 M-, and 1 C-class) with longitudes $\ge$65$^\circ$, and all flares below M5. Even with this subsample, our disagreements with \citet{Burtseva2013} and \citet{Petrie2010} concerning the positive and negative flux changes, and the flux dependence with $\mu$ remained. The correlation coefficients slightly decreased (e.g., from 0.60 to 0.57 in figure \ref{fig:areachanges}), but this is mainly because of the five X-class flares that were excluded due to their location. The differences between the studies that remain is the different spatial resolution and the solar cycle number. While the first should have no effect on the qualitative results, HMI has not yet observed two solar cycles to test the second difference. In summary, the reasons for the differences to previous studies are not because of the limb distances of the flares, the flare strengths, or the fitting method, but they remain unclear.}
The lack of a dependence on $\mu$ could indicate that loops do not simply increase or decrease. Instead, (un)twisting of loops may not have a preferred direction, and therefore no center-to-limb dependence may be expected.
 
\subsection{\blos{} and the Length of the Impulsive Phase}

We found that neither the duration of the impulsive phase nor the total duration of the flare were related to \blos{}. 
\citet{Sudol2005} and \citet{Petrie2010} reported the onset of \blos{} after the emission starts in the {\it GOES} thermal bandpass. The temporal parameters of the magnetic field change, such as its duration and its start and end times, are highly dependent on the quality of the fit of the stepwise function. Our manual verification showed that some fits are not reliable and therefore we do not to give statistics for these quantities. However, for the majority of  \blos{} ($\gtrsim$ 95\%), including the {double stepwise changes}, the onset of the changes occurred before the maximum of the {\it GOES} light curves (=during the impulsive phase). Some of them started slightly before the SXR emission increased (see examples in Figure~\ref{fig:classes} panels b, c, f, and g {and Figure~\ref{fig:allobs}}), which is earlier than other authors reported, { but could at least partially be explained with pre-flare events visible in SXR. High-resolution X-ray imaging spectroscopy would be required to clarify the timing of accelerated particles vs.~the magnetic field change}.

We observed double stepwise changes (figure~\ref{fig:classes}(l)) during the large flares X3.1 - \texttt{SOL2014-10-24T21:41}, X2.0 -- \texttt{SOL2014-10-27T14:47}, and X1.6 -- \texttt{SOL2014-10-22T 14:28}. These three events were located in the same AR12192, and they had long {\it GOES} durations of 66 min, 57 min, and 48 min, respectively. The {\it GOES} light curves during these events showed a second, superimposed flare before the {\it GOES} peak, and also before the second stepwise change. This may be an indication of a second magnetic reconnection process during these events. These results support that \blos{} are commonly observed during the impulsive phase of the flares \citep{Cliver2012}.

\subsection{\blos{}  and Flux Emergence/Cancellation}
We investigated how fast the magnetic field changes compared with regular flux emergence rates. We found that 98.1\% of \blos{} have a rate of change ranging from 1 to $\sim$300 \Mx{} min$^{-1}$. 
The magnetic flux emergence in the quiet Sun has a typical rate of $\sim$0.08 - 0.8 \Mx{} min$^{-1}$ \citep[e.g.,][]{Gosic2016,Smitha2016}. 
\citet{Parnell2009} found a power-law distribution of the rate of magnetic flux emergence, and \citet{Thornton2011} reported a global flux emergence rate that covers ephemeral regions {and} sunspots of about $\sim$450 \Mx{} day$^{-1}$ (0.3 \Mx{} min$^{-1}$). 

 Our data do not allow us to distinguish between actual flux emergence and geometric changes of the field. Therefore, we can only conclude that if actual flux emergence is occurring, it must be on much faster timescales than for the quiet Sun or for ARs.

The emergence of a tilted flux tube was suggested as a mechanism that may produce \blos{} \citep{Spirock2002, Wang2002a}. Our results are not consistent with this idea. The observed flux emergence rates for different solar features are orders of magnitudes lower than the rate of change observed during magnetic field changes. The flux emergence/cancellation during flares and the \blos{} might be unrelated \citep{Burtseva2013}. Therefore, we believe that regular flux emergence is probably not responsible for the magnetic field changes during flares.

\subsection{\blos{} and their Frequency Distribution}
We observed that the frequency distribution of the strength of \blos{} decreases exponentially ($dN/d$\blos{}, figure \ref{fig:statistics}). Changes up to 750 Mx cm$^{-2}$ were observed. We compared these results with previous studies by reconstructing their frequency distributions of \blos{}. 
The right panel in figure \ref{fig:statistics} shows the histograms of the frequency of \blos{} from \citet{Sudol2005} (purple), \citet{Petrie2010} (orange), an X-class flare from \citet{Burtseva2015} (yellow), and \citet{Kleint2017} (the HMI photospheric \blos{} in light green, and the chromospheric \blos{} in dark green). The distributions $dN/d$\blos{} from \citet{Sudol2005, Petrie2010} are the pixels that the authors considered best to represent their sample of flares.  
Several studies \citep{Sudol2005, Petrie2010, Burtseva2015} used data from GONG that has a plate scale of $\sim$2\farcs5 pixel$^{-1}$ ($\sim5$ times larger than HMI), which could explain the overall lower number of detected changes. \citet{Kleint2017} recorded chromospheric \blos{} with the IBIS instrument in the \ion{Ca}{2} 8542 \AA{} line with a 
binning of 0\farcs2 pixel$^{-1}$ ($\sim2.5$ times smaller than HMI). \citet{Kleint2017} noted that their number of changes below 200 \Mx{} are underestimated because of their S/N, therefore we omitted points below 200 \Mx{} from our fitting.

We observed that the frequency distributions of \blos{} from previous studies also followed exponential distributions. While the exponent is different for each flare, our results (left panel) and all other reports (right panel) lead to the conclusion that the exponential nature of the frequency distribution of \blos{} does not depend on the spatial scale of the instrument that is used. Table \ref{tab:exponents} summarizes the decay exponents of the frequency distribution of \blos{}. The scatter of the exponents is large, and they do not show any dependence on the flare magnitude. {They may also slightly differ for a given flare if different time ranges are analyzed, because there is no unique criterion for the duration of flare-related changes and some observed changes may be interpreted differently in different time ranges.}

\begin{deluxetable}{lcc}
\tabletypesize{\footnotesize }
\tablecolumns{3} 
\tablecaption{{Decay exponents of the frequency distribution of the sizes of \blos{} for different flares.} \label{tab:exponents}}
\tablehead{\colhead{Flare} & \colhead{Exponent (\Mx{})} & \colhead{Note}}
\startdata
{All flares} & ${86.6\pm27.7}$\tablenotemark{a} & This study\\
X-class & ${85.7\pm 19.8}$\tablenotemark{a} & 18 X \\
M-class  & ${89.3\pm 31.7}$\tablenotemark{a} & 35 M\\
C-class  & ${74.5\pm 16.3}$\tablenotemark{a} & 6 C\\\\
\citet{Sudol2005}\tablenotemark{b} & 108.4 & {15 X} \\
\citet{Petrie2010}\tablenotemark{b}  & 101.8 & 38 X, 39 M \\
\citet{Burtseva2015}\tablenotemark{c} &28.3, 34.8, 38.5, 100.9& 4 X \\
\citet{Kleint2017}  & {42.9} (phot), {79.3} (chro) & {X1}\\
\enddata
\tablenotetext{a}{{One standard deviation ($\sigma$).}}
\tablenotetext{b}{Distribution of selected pixels that best represent the flare sample.}
\tablenotetext{c}{Flares X10.0, X8.3, X6.5, and X2.6. {Their X1.0 flare only} showed 13 \blos{}, therefore we did not reconstruct the frequency distribution for this event.}
\end{deluxetable}

\section{Summary and Conclusions} \label{sec:discussion}

We presented a survey of \blos{} during 75 solar flares using {\it SDO}/HMI data, which includes B- to X-class flares distributed from disk center to the limb. We address key results from our study:
\begin{itemize}
\item  Energetic flares are more likely to show permanent changes. All flares with {\it GOES} classes larger than M1.6 showed \blos{} when using HMI data. 
\item  6 of 17 of the C-class flares were associated with \blos{} larger than 80 \Mx{}. A C3.0 flare was the smallest flare where we detected \blos{}. 
\item We found that the peak {\it GOES} X-ray flux of the flares is well correlated with the total signed/unsigned magnetic flux changes as well as with the area perturbed by the \blos{}. {This may be explained as larger flares having more energy available to change the magnetic field potentially via the Lorentz force.} 
\item The frequency distribution of the number of \blos{} and the distance of \blos{} with respect to the PIL decreases exponentially. This may imply a sheared PIL next to which \blos{} are observed.

\item The frequency distribution of the sizes of \blos{} decreases exponentially ($dN/d$\blos{}$\sim e^{-|\BLOS{}|/{88.4}}$). We compared our findings with previous observations made with different instruments. We conclude that the exponential behavior of the $dN/d$\blos{} does not depend on the resolution of the instrument or on types of flares.
\end{itemize}

The origin of \blos{} is still unclear. Therefore, we will investigate in future studies if they are spatially and temporally related to different flare phenomena, for example, white-light (WL) emission, non-thermal X-ray emission, and sunquakes \citep[acoustic waves seen in the photosphere during flares, e.g.,][]{Kosovichev1998}. \citet{Burtseva2015} noticed that the location of the \blos{} coincided with hard X-ray emission in the early phase of flares, but their statistics were limited.
However, it is still unclear if accelerated particles are related to the production of \blos{}, or if they merely coincide as separate flare phenomena. Our flares were selected to include RHESSI coverage, which will enable a direct comparison with particle acceleration sites. 

Sunquakes are another solar mystery. They are only observed in some flares, and not even in all X-flares, and their physical origin is still debated. 
While different processes, such as the Lorentz force, non-thermal particles, hydrodynamic processes, or back-warming radiation were proposed as their drivers \citep{Machado89,Zharkova93,Allred05,Donea2005,Hudson2008}, the 
responsible mechanism is still under debate \citep[see e.g.,][]{Judge2014ApJ796}. By utilizing {\it SDO}'s new data product of full-vector data at a 90 s cadence \citep{Sun2017}, we plan to map vector changes and to calculate the changes of the Lorentz force and compare them to occurrence and position of sunquakes. Future studies with new instruments, especially chromospheric polarimetry, will shed light on the 3D structure of magnetic field changes, ultimately allowing us to determine how flare energy is dissipated.
\bigskip

J.~S.~Castellanos~Dur\'an was funded by the grant UNAL--HERMES 26675, {\it J\'ovenes Investigadores}, COLCIENCIAS -- 645, Colombia. {We thank the referee for their careful reading and suggestions.}
{\it SDO} is a mission for NASA's Living With a Star program.

\facility{{\it SDO}/HMI, {\it SDO}/AIA.}

\appendix\label{sec:appendix}

{{Here, we describe why we sometimes adapted the step size calculated by equation \ref{eq:sudol}, which}
is given by $\BLOS{}=2c$. If the slope of the background and the duration of the step increase, the value obtained for the amplitude of the step {depends on assumptions whether the slope remains constant or changes during the flare. While some previous studies \citep[e.g.][]{Sudol2005,Petrie2010} assumed a constant slope throughout the flare, thus using the term $a+bt$ for a linear background evolution, we decided to determine the effective step size without assuming that the slope continues linearly throughout the flare. Our reasoning was that we have seen changes in the flux emergence/cancellation rate at flare time and we wanted to estimate the step size more conservatively}. For this reason we introduced two additional methods to estimate \blos{} in section \ref{sec:locationblos}. An example of a common stepwise profile is shown in figure \ref{appendixfig}. The magnetic field evolution (black diamonds) is fitted well by equation \ref{eq:sudol} (red line). The fit parameters are shown in the figure's title, where $c_{fit}=-231$ \Mx{}, the slope of the background $b_{fit}=3.3$ \Mx{} min$^{-1}$, and the duration of the change $n^{-1}_{fit}=8.3$ min. In this case, the amplitude of the step calculated as $2c$ is -463 \Mx{}, which is two times larger {than what our method derives. This also implies that former publications may have obtained different, possibly larger step sizes {and that our method is slightly biased toward small $2c$}. 
}
 \begin{figure}[hptb]
 \begin{center}
 \includegraphics[width=.49\textwidth]{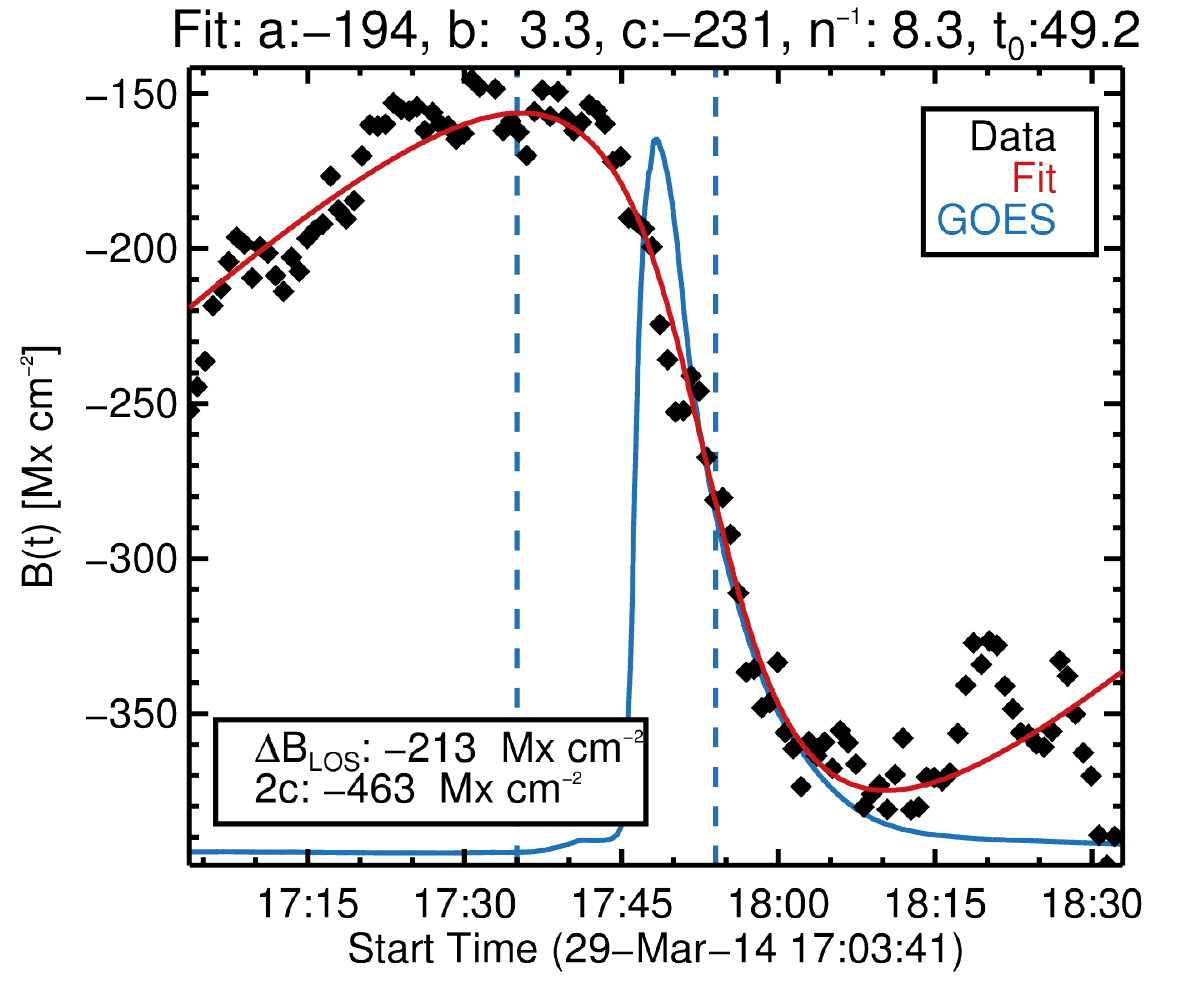}
 \caption{{Typical stepwise change (black diamonds) that is well fitted by equation \ref{eq:sudol} (red line). The blue line is the {\it GOES} light curve, and the dashed lines represent the start and end {\it GOES} times (from the flare database). $2c$ is -463 \Mx{}, while our method found a step size of -213 \Mx{}.\label{appendixfig} }}
 \end{center}
 \end{figure}

\bibliographystyle{aasjournal}
\bibliography{references}

\end{document}